\documentclass[a4paper,11pt]{article}
\pdfoutput=1
\usepackage{jheppub}
\usepackage{amsmath,amssymb,bm,graphicx,epsf,colordvi}
\usepackage{slashed}
\def\bea #1\eea{\begin{align} #1\end{align}} 
\usepackage{diagbox}
\usepackage{lipsum}
\usepackage{soul}
\usepackage{braket}
\usepackage{bm}
\allowdisplaybreaks 
\usepackage{color}
\usepackage[dvipsnames]{xcolor}

\usepackage{mathrsfs}
\usepackage{appendix}
\usepackage{bm}
\usepackage{lipsum}

\newcommand{\bef}{\begin{figure}[hbt]\centering}
\newcommand{\eef}{\end{figure}}

\definecolor{darkgreen}{rgb}{0.13,0.55,0.13}
\usepackage{mathtools}
\usepackage{subfigure}
\usepackage{booktabs}
\usepackage{graphicx}
\graphicspath{ {./figure/} }
\usepackage{caption}
\usepackage{lipsum}

\newcommand{\nnu}{\nonumber\\}

\newcommand{\beq}{\begin{equation}}
\newcommand{\eeq}{\end{equation}}

\def \be  {\begin{equation}}
\def \ee  {\end{equation}}
\def \ba  {\begin{eqnarray}}
\def \ea  {\end{eqnarray}}

\newcommand{\jet}{{J}}
\newcommand{\approptoinn}[2]{\mathrel{\vcenter{
  \offinterlineskip\halign{\hfil$##$\cr
    #1\propto\cr\noalign{\kern2pt}#1\sim\cr\noalign{\kern-2pt}}}}}

\allowdisplaybreaks

\usepackage{graphicx}

\usepackage{stackengine}

\usepackage{atbegshi,picture}
\AtBeginShipoutNext{\AtBeginShipoutUpperLeft{%
\put(\dimexpr\paperwidth-0.5cm\relax,-5cm){\makebox[0pt][r]{MIT-CTP/5632}}%
}}
\title{Probing Transverse Momentum Dependent Structures with Azimuthal Dependence of Energy Correlators}

\author[a,b,c]{Zhong-Bo Kang}
\author[d,e]{, Kyle Lee}
\author[f,g]{, Ding Yu Shao}
\author[e,a]{, Fanyi Zhao}

\affiliation[a]{Department of Physics and Astronomy, University of California, Los Angeles, CA 90095, USA}
\affiliation[b]{Mani L. Bhaumik Institute for Theoretical Physics, University of California, Los Angeles, CA 90095, USA}
\affiliation[c]{Center for Frontiers in Nuclear Science, Stony Brook University, Stony Brook, NY 11794, USA}
\affiliation[d]{Nuclear Science Division, Lawrence Berkeley National Laboratory, Berkeley, California 94720, USA}
\affiliation[e]{Center for Theoretical Physics, Massachusetts Institute of Technology, Cambridge, MA 02139}
\affiliation[f]{Department of Physics, Center for Field Theory and Particle Physics and Key Laboratory of Nuclear Physics and Ion-beam Application (MOE), Fudan University, Shanghai 200433, China}
\affiliation[g]{Shanghai Research Center for Theoretical Nuclear Physics, NSFC and Fudan University, Shanghai 200438, China}

\emailAdd{zkang@ucla.edu}
\emailAdd{kylel@mit.edu}
\emailAdd{dingyu.shao@cern.ch}
\emailAdd{fanyi@mit.edu}

\abstract{We study the azimuthal angle dependence of the energy-energy correlators $\langle \mathcal{E}(\hat{n}_1)\mathcal{E}(\hat{n}_2)\rangle$ in the back-to-back region for $e^+e^-$ annihilation and deep inelastic scattering (DIS) processes with general polarization of the proton beam. We demonstrate that the polarization information of the beam and the underlying partons from the hard scattering is propagated into the azimuthal angle dependence of the energy-energy correlators. In the process, we define the Collins-type EEC jet functions and introduce a new EEC observable using the lab-frame angles in the DIS process. Furthermore, we extend our formalism to explore the two-point energy correlation between hadrons with different quantum numbers $\mathbb{S}_i$ in the back-to-back limit $\langle \mathcal{E}_{\mathbb{S}_1}(\hat{n}_1)\mathcal{E}_{\mathbb{S}_2}(\hat{n}_2)\rangle$. We find that in the Operator Product Expansion (OPE) region the nonperturbative information is entirely encapsulated by a single number. Using our formalism, we present several phenomenological studies that showcase how energy correlators can be used to probe transverse momentum dependent structures.}

\begin{document}

\maketitle

\section{Introduction}\label{sec:intro}
The energy-energy correlator (EEC)~\cite{Basham:1978bw,Basham:1978zq} is one of the earliest infrared and collinear (IRC) safe observables~\cite{Kinoshita:1962ur,Lee:1964is}. It measures the energy correlations as a function of the angle $\theta$ between two detectors. This allows EEC to be represented in terms of the correlation function of energy flow operators~\cite{Sveshnikov:1995vi,Tkachov:1995kk,Korchemsky:1999kt,Bauer:2008dt,Hofman:2008ar,Belitsky:2013xxa,Belitsky:2013bja,Kravchuk:2018htv} defined as
\begin{align}
\label{eq:energyflow}
\mathcal{E}(\vec{n})=\int_0^{\infty} d t \lim _{r \rightarrow \infty} r^2 n^i T_{0 i}(t, r \vec{n})\,,
\end{align}
making them a subject of extensive study in conformally invariant $\mathcal{N}=4$ super-Yang-Mills (SYM) theory~\cite{Hofman:2008ar,Belitsky:2013xxa,Belitsky:2013bja,Belitsky:2013ofa,Chicherin:2023gxt}. The EEC as event shape
observables has been extensively measured in $e^+e^-$ collisions~\cite{SLD:1994idb,L3:1992btq,OPAL:1991uui,TOPAZ:1989yod,TASSO:1987mcs,JADE:1984taa,Fernandez:1984db,Wood:1987uf,CELLO:1982rca,PLUTO:1985yzc,OPAL:1990reb,ALEPH:1990vew,L3:1991qlf,SLD:1994yoe} and most recently has been a target of study in electron-hadron scatterings for the future Electron-Ion Collider (EIC)~\cite{Li:2021txc,Neill:2022lqx}. In hadron-hadron collisions, a generalization of the EEC called transverse EEC (TEEC) has been measured and studied at the Large Hadron Collider (LHC) kinematics~\cite{Gao:2019ojf,ATLAS:2015yaa,ATLAS:2017qir,ATLAS:2020mee,Alvarez:2023fhi}. 
The study of EEC has played a crucial role in advancing our understanding of fundamental particles and their interactions. For example, these measurements of EEC provide one of the sharpest determination of the strong coupling constant for both hadron and lepton collider environments~\cite{ParticleDataGroup:2020ssz,deFlorian:2004mp,Burrows:1995vt,Kardos:2018kqj,Alvarez:2023fhi}. 

Over the last several decades, with particularly reinvigorated efforts in recent years, there have been significant advances in the theoretical studies of the EEC. There have been rigorous computations at fixed orders for arbitrary $\theta$ angles~\cite{Schneider:1983iu, Falck:1988gb, Glover:1994vz, Kramer:1996qr, Ali:1982ub, Ali:1984gzn, Richards:1982te, Richards:1983sr, Catani:1996jh, Tulipant:2017ybb,Belitsky:2013xxa,Belitsky:2013bja}, where the state-of-the-art computations provide an analytical expressions up to NLO~\cite{Dixon:2018qgp,Luo:2019nig} and numerical results at NNLO accuracy~\cite{DelDuca:2016ily} in QCD, and analytical evaluations at both NLO~\cite{Belitsky:2013ofa} and NNLO~\cite{Henn:2019gkr} for $\mathcal{N}=4$ SYM. There has also been many important works to better understand the nonperturbative structure of the EEC~\cite{Schindler:2023cww,Nason:1995np,Korchemsky:1999kt,Belitsky:2001ij}. Furthermore, the factorization structure of the singular regions of the EEC in the back-to-back 
 ($\theta\to \pi$) ~\cite{Kodaira:1981nh, Kodaira:1982az, deFlorian:2004mp, Tulipant:2017ybb, Moult:2018jzp, Ebert:2020sfi,Duhr:2022yyp, Moult:2019vou,Collins:1981uk} and collinear limits ($\theta\to 0$)~\cite{Dixon:2019uzg,Korchemsky:2019nzm,Konishi:1978yx,Konishi:1978ax} have been better understood in recent years using the soft collinear effective field theory (SCET)~\cite{Bauer:2001yt,Bauer:2000yr,Bauer:2001ct}, allowing them to be studied to unprecedented accuracy in both regions.\footnote{The collinear limit has been computed at NNLL accuracy. For rich interplay and wide-ranging applications of EEC in the collinear limit with the jet substructure program, see~\cite{Lee:2023npz,Devereaux:2023vjz,Craft:2022kdo,Lee:2022ige,Yang:2023dwc,Andres:2022ovj,Procura:2022fid,Chen:2020vvp}.} In particular, connection to the Transverse-Momentum Dependent (TMD) observables~\cite{Boussarie:2023izj} in the back-to-back limit has been observed and studied in the context of unpolarized processes in both $e^+e^-$ and appropriately modified definition of the EEC for the $ep$ collisions. With the state-of-the-art computation of the $4$-loop rapidity anomalous dimension~\cite{Moult:2022xzt,Duhr:2022yyp}, the EEC in the back-to-back limit has been calculated at very impressive N$^4$LL accuracy~\cite{Duhr:2022yyp}. Within the $\mathcal{N}=4$ SYM, even the subleading power resummation in the back-to-back limit has been carried out~\cite{Chen:2023wah} using techniques from conformal bootstrap. Many developments of EEC like this in the context of $\mathcal{N}=4$ SYM give hope of importing valuable techniques from the CFTs to improve and better our understanding for QCD as well.

\begin{figure}[h]
\centering
\includegraphics[width=0.7\textwidth]{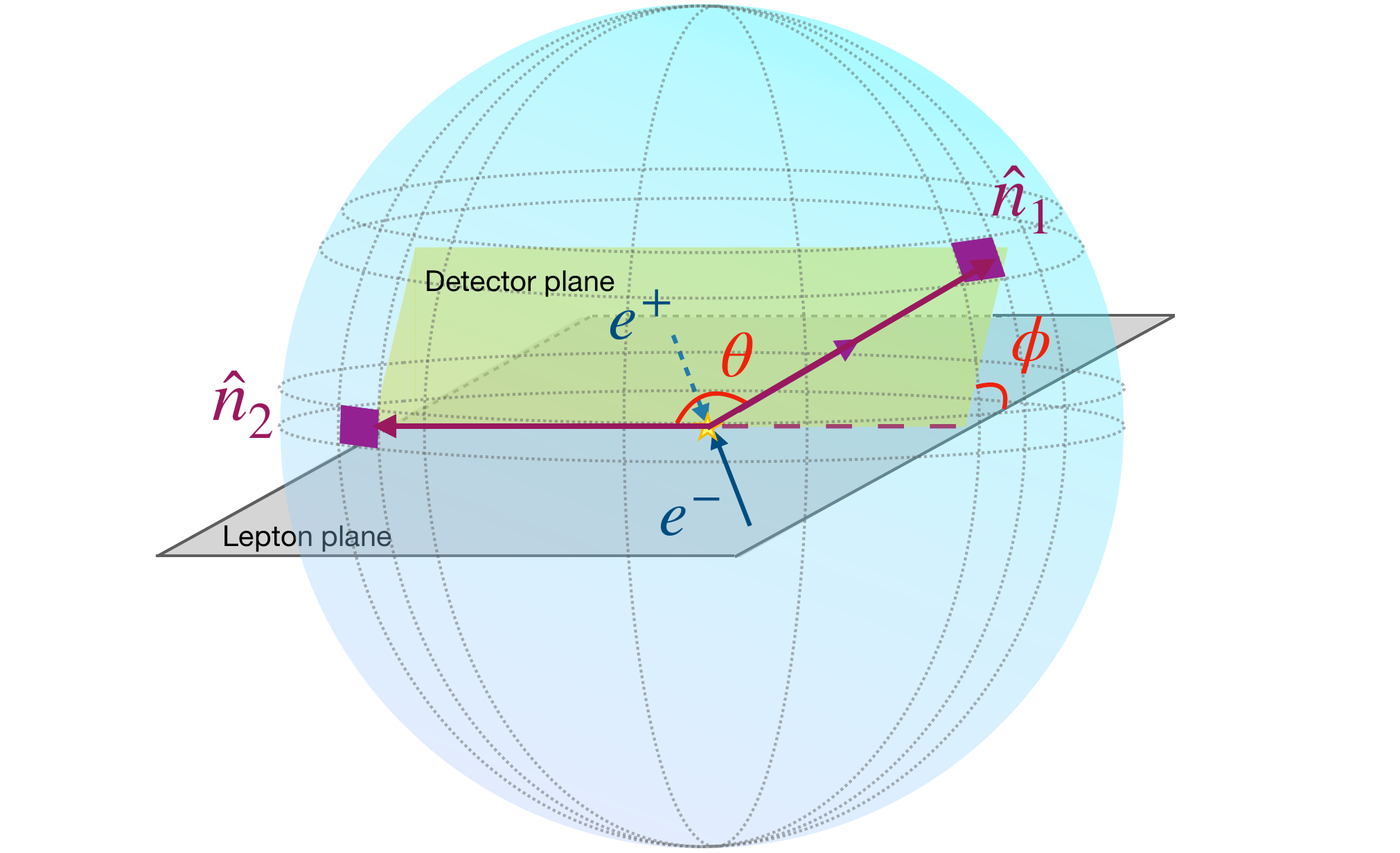}
\caption{Illustration of {{EEC}} for $e^+e^-$ annihilation. Here $\theta$ is the angular separation between the two detectors pointing along $\hat{n}_1$ and $\hat{n}_2$ and $\phi$ is the azimuthal angle between the plane formed by these detectors (referred to as ``detector plane'', shown as yellow) and the plane generated by the direction $\hat{n}_2$ and the beam (referred to as ``lepton plane'', shown as gray).}
\label{fig:intro}
\end{figure}

While significant progress has been made for the EEC, almost all the existing work concentrate on studying the EEC as a function of the angle $\theta$ only.\footnote{Refer to~\cite{Chen:2020adz,Barata:2023zqg,Li:2023gkh} for several important recent examples that are exceptions, notably considering the azimuthal angle dependence in the collinear limit of the EEC, as well as for the nucleon EEC.} However, as shown already in the original paper where the EEC was first introduced~\cite{Basham:1978bw,Basham:1978zq}, when considering the incoming beam direction in $e^+e^-$ collisions, there would be two additional independent angles one has to introduce in order to fully describe the EEC. As shown in Fig.~\ref{fig:intro}, in addition to the usual angle $\theta$ between the two detectors, one also has an azimuthal angle difference $\phi$ between the plane formed by the detectors (referred to as ``detector plane'', shown as yellow) and the plane generated by $\hat{n}_2$ and the beam (referred to as ``lepton plane'', shown as gray). The very first proposal of energy correlators in~\cite{Basham:1978bw} included an LO calculation of the EEC as a function of the angle $\theta$ and the azimuthal angle $\phi$. However, at the end of the paper, the authors decided to integrate over the azimuthal angle $\phi$, indicating that the importance of this azimuthal dependence in the EEC was not realized back then. Since then, higher-order computations of energy correlators have consistently integrated such azimuthal dependence.

In recent years, spin asymmetries have been widely studied in the hadron physics community. Because these asymmetries provide nontrivial quantum correlations and imaging for the hadron structure, their studies have been one of the most important scientific thrusts for the future EIC~\cite{Accardi:2012qut,AbdulKhalek:2021gbh,AbdulKhalek:2022hcn}. Motivated by the fact that many spin asymmetries arise from the azimuthal correlations, we study the {\it azimuthal dependent} EEC. We will develop the theoretical formalism to demonstrate the full potential of such azimuthal dependent EEC observables, in particular in terms of exploring the nucleon structure.

In this paper we concentrate on the EEC in the back-to-back region where the angle between the detectors approaches $\theta \to \pi$ and study its azimuthal angle dependence. For convenience, we also often use the variable $\tau$ defined as\footnote{Note that $z$ variable defined as $z = 1-\tau$ is also often used in the literature~\cite{Moult:2018jzp,Ebert:2020sfi}.}
\begin{align}
\label{eq:taudef}
    \tau = \frac{1+\cos \theta}{2}\,,
\end{align}
and thus the back-to-back region corresponds to $\tau\to 0$. Then the azimuthal dependent EEC in the $e^+e^-$ collisions is defined as
\begin{align}
\label{eq:EECazdef}
\mathrm{{{EEC}}}_{e^+e^-}(\tau,\phi) \equiv\frac{d \Sigma_{e^+e^-}}{d \tau d\phi}=\frac{1}{2}\sum_{1, 2} \int d \sigma z_1 z_2 \,\delta\left(\tau - \frac{1+\cos \theta_{12}}{2}\right) \delta(\phi - \phi_{12})\,,
\end{align}
where the weighted sum over pairs of hadrons produced in the final state give correlation between the energy flown into the detectors in the form of hadrons detected. We also use $z_{1,2} = 2E_{1,2}/Q$ to denote the energy fractions of the final-state hadrons $1,2$ flown into the detectors $1,2$, respectively. This definition of the EEC can also be recasted using the operator definition of the energy flow given in Eq.~\eqref{eq:energyflow} as 
\begin{align}
\frac{1}{\sigma_{\rm tot}} \frac{d \Sigma_{\rm e^+e^-}}{d \tau d\phi}=\frac{\left\langle\mathcal{O}\mathcal{E}\left(\vec{n}_1\right) \mathcal{E}\left(\vec{n}_2\right) \mathcal{O}^{\dagger}\right\rangle}{\left\langle\mathcal{O O}^{\dagger}\right\rangle}\,,
\end{align}
where $\mathcal{O}$ is a source operator that creates the excitation that is detected in the form of energy carried by the hadrons in the asymptotic detector.

Below, we will demonstrate the similarity between the azimuthal dependent EEC and the usual TMD factorization formalism. In particular, we show in the back-to-back region that a new term $\propto \cos(2\phi)$ will arise for $\mathrm{{{EEC}}}_{e^+e^-}(\tau,\phi)$, which is related to the Collins fragmentation function~\cite{Collins:1992kk}, one of the most extensively discussed polarized transverse momentum dependent functions (TMDs), describing the fragmentation of a transversely polarized quark into an unpolarized hadron. This fragmentation process correlates the transverse momentum of the outgoing hadron with the transverse polarization of the quark, giving rise to non-trivial azimuthal angular asymmetries. For the $e^+e^-$ annihilation, two transversely polarized quarks can be produced from the unpolarized $e^+e^-$ without violating spin conservation. The Collins asymmetries then manifest itself with the azimuthal asymmetry $\cos (2\phi)$ in the azimuthal dependent EEC observable. 

In addition, we introduce a similar version of azimuthal dependent EEC in deep inelastic $ep$ scattering (DIS), extending the unpolarized case considered in~\cite{Li:2021txc} using the Breit frame. We show that the azimuthal dependence in the EEC would allow us to probe the transverse momentum dependent parton distribution functions (TMD PDFs). In particular, in the context of $ep$ collisions employing polarized electron and/or proton beams, the EEC in DIS in the back-to-back region exhibits a remarkable correlation between the EEC jet functions defined below and the polarized and unpolarized TMD PDFs, highlighting the immense potential of azimuthal dependent {{EEC}} as a novel tool for probing TMD PDFs and advancing our knowledge of the internal structure of nucleons. Furthermore, building on the insights presented in~\cite{Gao:2022bzi}, we introduce a new EEC observable in DIS defined using the lab-frame angle $q_*$. By solely relying on angles defined within the Lab frame, this observable offers an order-of-magnitude improvement in the anticipated experimental resolution at the EIC.

The rest of the paper is structured as follows. Section~\ref{sec:kinematics} presents a comprehensive analysis of the factorization framework governing azimuthal dependent energy-energy correlators in $e^+e^-$ and $ep$ collisions. We introduce the Collins-type EEC jet function, which has a close relation to the Collins fragmentation function. In Section~\ref{sec:jetfunc}, we study the properties of the unpolarized and Collins-type EEC jet functions. Section~\ref{sec:pheno} presents our phenomenological study to demonstrate the potential of azimuthal dependent EEC observables for probing nucleon structures. Finally, we conclude our work in Section~\ref{sec:conclusion}.

\section{Factorization Formalism for Azimuthal Dependent EEC}\label{sec:kinematics}
In this section, we study the azimuthal dependent {{EEC}} observables for both $e^+e^-$ annihilation and DIS processes. We show that in the back-to-back region, they can be related to the TMD factorization framework. We demonstrate that in the factorization formalism, beside the usual unpolarized EEC jet function which is related to the unpolarized TMD fragmentation functions, the Collins-type EEC jet function arises that is closely connected with the Collins fragmentation functions. We write down all the azimuthal angle dependent correlations for the {{{EEC}}}$(\tau, \phi)$ observables for both $e^+e^-$ and DIS processes. In our study of the EEC in DIS, we extend the version adapted from the Breit frame as cited in~\cite{Li:2021txc}. Drawing inspiration from~\cite{Gao:2022bzi}, we also introduce a new EEC utilizing lab-frame angles. Furthermore, we demonstrate the applicability of EEC in DIS in probing both polarized and unpolarized nucleon structures.

\subsection{Azimuthal dependent EEC for $e^+e^-$ annihilation}
We will first discuss the azimuthal dependent {{EEC}} for the $e^+e^-$ annihilation. Let us start with specifying the details of the coordinate frame in which the observables will be measured, shown already in Fig.~\ref{fig:intro}. As often done in the TMD factorization for $e^+e^-$, we adopt the so-called Gottfried-Jackson (GJ) frame~\cite{Gottfried:1964nx,Pitonyak:2013dsu} by aligning the detector 2 along the $\hat{n}_2=\hat{n}_z$. 
Therefore, the angle $\theta$ between the two detectors is the polar angle of the detector 1 in the GJ frame. With this setup, we can easily measure the azimuthal angle of the detector plane with respect to the lepton plane, as mentioned already in the Introduction and shown clearly in Fig.~\ref{fig:intro}. This coordinate frame is slightly different from what was introduced in the original EEC paper~\cite{Basham:1978bw} where the $+z$ axis is defined to be along the incoming lepton $e^-$ beam, while the azimuthal angle $\phi$ is kept the same. In general, one can also parameterize the difference between the two frame choices by keeping track of the angle between the $e^-$ beam and the detector $2$. However, as we integrate over this angle and only measure $\theta$ and $\phi$, the two coordinate frame choice give the same results. This demonstrates that our choice of using GJ frame was not unique, though it is useful to visualize the measurements. 

In the back-to-back limit of the two detectors where the angle approaches $\theta \to \pi$ (or $\tau \to 0$), we find up to power corrections,
\begin{align}
\tau = \frac{\bm{P}_{h_1\perp}^2}{z_1^2Q^2}\,,
\end{align}
for a given pair of hadrons in the two detectors. Here, $\bm{P}_{h_1\perp}$ is the transverse momentum of the hadron $h_1$ in the GJ frame, namely with respect to $\hat{n}_2$ in the pair production. The azimuthal angle associated with $\bm{P}_{h_1\perp}$ is simply $\phi$ in the GJ frame, as shown in Fig.~\ref{fig:intro}. The energy fractions $z_{i = 1,2}$ are given by
\begin{align}
    z_i = \frac{2P_{hi}\cdot q}{Q^2} = \frac{2E_i}{Q}\, ,
\end{align}
where $q = \ell_{e^+} + \ell_{e^-}$ with $Q^2 = q^2$. For convenience, we also introduce $\bm{q}_T = -\bm{P}_{h_1\perp}/z_1$. 

With the relation between $\tau$ and ${\bm q}_T$ at hand in the back-to-back region, the azimuthal dependent {{EEC}} for $e^+e^-$ defined in Eq.~\eqref{eq:EECazdef} can then be related to the $\bm{q}_T$-differential cross section as
\begin{align}
\label{eq:EECazdef2}
\mathrm{{{EEC}}}_{e^+e^-}(\tau,\phi) 
&=  \frac{d\Sigma_{e^+e^-}}{d\tau d\phi} \nnu
&=\frac{1}{2} \sum_{1,2}\int d^2\bm{q}_T\, dz_1\, dz_2 \, z_1 \, z_2 \,\frac{d\sigma}{dz_1 dz_2 d^2\bm{q}_T} \,\delta\left(\tau - \frac{\bm{q}_T^2}{Q^2}\right) \delta(\phi - \phi_{12})\,,
\end{align}
where the standard TMD factorization for the back-to-back di-hadron process in $e^+e^-$ is given by~\cite{Boer:2008fr,Pitonyak:2013dsu}
\begin{align}
\frac{d\sigma}{dz_idz_j d^2{\bm q}_T}=&\,\sigma_0\,H(Q,\mu)\sum_q e_q^2\int d^2{\bm p}_{1\perp}d^2{\bm p}_{2\perp}d^2{\bm \lambda}_{\perp}\delta^2\left(\frac{{\bm p}_{1\perp}}{z_1}+\frac{{\bm p}_{2\perp}}{z_2}-{\bm \lambda}_{\perp}+{\bm q}_T\right){S}({\bm \lambda}_{\perp}^2,\mu,\nu)\nnu
&\hspace{-1.2cm}\times\bigg[D_{1,h_1/q}^{(u)}(z_1,{\bm p}_{1\perp}^2,\mu,\zeta/\nu^2)D^{(u)}_{1,h_2/\bar{q}}(z_2,{\bm p}_{2\perp}^2,\mu,\zeta/\nu^2)+\cos (2\phi_{12})
\left({\bm \hat{\bm q}}_{T,\alpha}{\hat{\bm q}}_{T,\beta} - \frac{1}{2}g_{\perp,{\alpha\beta}}\right)
\nnu
&\hspace{-0.6cm}\times\frac{\bm p_{1\perp}^\alpha}{z_1M_1}H_{1,h_1/q}^{\perp(u)}(z_1,{\bm p}_{1\perp}^2,\mu,\zeta/\nu^2)\frac{\bm p_{2\perp}^\beta}{z_2M_2}H_{1,h_2/\bar{q}}^{\perp(u)}(z_2,{\bm p}_{2\perp}^2,\mu,\zeta/\nu^2)\bigg]\,.
\label{eq:diffcskt}
\end{align}
Here we have integrated over the angle between the hadron $h_2$ and the incoming beam direction from the conventional TMD factorization expression. It is natural to concentrate on the dependence on the angle $\theta$ and $\phi$ only from the EEC perspective as it is an inclusive measurement over all the hadron pairs as a function of angular separation of detectors in which they are found. 
The born cross-section is given by $\sigma_0 = 4\pi N_c\alpha_{\rm em}^2/3Q^2$. On the other hand, $D_{1,h/q}^{(u)}(z,{\bm p}_{\perp}^2,\mu,\zeta/\nu^2)$ and $H_{1,h/q}^{\perp(u)}(z,{\bm p}_{\perp}^2,\mu,\zeta/\nu^2)$ are the unsubtracted ($u$) unpolarized and Collins transverse momentum dependent fragmentation functions (TMD FFs)~\cite{Boussarie:2023izj}, with $\bm{p}_{\perp}$ the hadron transverse momentum with respect to the fragmenting parton. As we have mentioned already, Collins fragmentation function describes the process where a transversely polarized quark fragments into an unpolarized hadron with the quark transverse spin correlated with the hadron transverse momentum. Here $\mu$ and $\nu$ are the usual renormalization and rapidity renormalization scales, and $\zeta$ is the Collins-Soper scale, for details see Ref.~\cite{Boussarie:2023izj}. We also have $H(Q, \mu)$, the hard function, which at the next-to-leading order is (see e.g.~\cite{Becher:2008cf})
\begin{align}
    H(Q, \mu) = 1 + \frac{\alpha_s}{2\pi} C_F
    \left[3\ln\frac{Q^2}{\mu^2} - \ln^2\frac{Q^2}{\mu^2} - 8 + \frac{7\pi^2}{6}\right]\,.
\end{align}
Finally, ${S}({\bm \lambda}_{\perp}^2,\mu,\nu)$ is the soft function with $\bm{\lambda}_{\perp}$ describing the soft recoil of the fragmenting partons from being exactly back-to-back. The TMD factorization formalism can be transformed into the coordinate $b$-space as
\begin{align}
\frac{d\sigma}{dz_1dz_2 d^2{\bm q}_T}=&\,\sigma_0\,H(Q,\mu) \sum_q e_q^2\int\frac{d^2{\bm b}}{(2\pi)^2}e^{-i{\bm b}\cdot{\bm q}_T}S({\bm b}^2,\mu,\nu)
\nnu
&\times\bigg[\tilde{D}^{(u)}_{1,h_1/q}(z_1,b,\mu,\zeta/\nu^2)\tilde{D}^{(u)}_{1,h_2/q}(z_2,b,\mu,\zeta/\nu^2)
+ \cos (2\phi_{12})
\left({\bm \hat{\bm q}}_{T,\alpha}{\hat{\bm q}}_{T,\beta} - \frac{1}{2}g_{\perp,{\alpha\beta}}\right)
\nnu
&\hspace{0.5cm}\times \tilde{H}_{1,h_1/{q}}^{\perp(u)\,\alpha}(z_1,{\bm b},\mu,\zeta/\nu^2)\tilde{H}_{1,h_2/{q}}^{\perp(u)\,\beta}(z_2,{\bm b},\mu,\zeta/\nu^2)\bigg]\,, 
\label{eq:diffcsb}
\end{align}
where the TMD FFs in the $b$-space are defined as
\begin{align}
\tilde{D}^{(u)}_{1,h/{q}}(z,b,\mu,\zeta/\nu^2)=&\int d^2{\bm p}_{\perp}e^{-i{\bm b}\cdot{{\bm p}_{\perp}}/{z}}D^{(u)}_{1,h/q}(z,{\bm p}_{\perp}^2,\mu,\zeta/\nu^2)\,,\\
\tilde{H}_{1,h/{q}}^{\perp(u)\,\alpha}(z,{\bm b},\mu,\zeta/\nu^2)=&\int d^2{\bm p}_{\perp}e^{-i{\bm b}\cdot{{\bm p}_{\perp}}/{z}}\frac{\bm p_{\perp}^\alpha}{zM} H_{1,h/q}^{\perp(u)}(z,{\bm p}_{\perp}^2,\mu,\zeta/\nu^2)\,,
\\
\equiv& \left(-\frac{i{\bm b}^\alpha}{2}\right) \tilde{H}_{1,h/{q}}^{\perp(u)}(z,b,\mu,\zeta)
\label{eq:Hperp(u)}
\end{align}
and likewise for the soft function ${S({\bm b}^2,\mu,\nu)}$ in the $b$-space, which has been computed to the three-loop order in~\cite{Li:2016ctv}. To simplify the notation, here we introduce the ``subtracted" TMD FFs as,
\begin{align}
\label{eq:subTMD FFs}
\tilde{F}(z,{\bm b},\mu,\zeta)=\tilde{F}^{(u)}(z,{\bm b},\mu,\zeta/\nu^2)\sqrt{S({\bm b}^2,\mu,\nu)}\,,
\end{align}
where $\tilde{F}$ can be unpolarized or the Collins TMD FFs. As a result, Eq.~\eqref{eq:diffcsb} is further written as
\begin{align}
\frac{d\sigma}{dz_1dz_2 d^2{\bm q}_T}=&\,\sigma_0\,H(Q,\mu) \sum_q e_q^2\int\frac{d^2{\bm b}}{(2\pi)^2}e^{-i{\bm b}\cdot{\bm q}_T}\bigg[\tilde{D}_{1,h_1/q}(z_1,b,\mu,\zeta)\tilde{D}_{1,h_2/q}(z_2,b,\mu,\zeta)
\nnu
&\hspace{-12mm}+ \cos (2\phi_{12})
\left({\bm \hat{\bm q}}_{T,\alpha}{\hat{\bm q}}_{T,\beta} - \frac{1}{2}g_{\perp,{\alpha\beta}}\right) \tilde{H}_{1,h_1/{q}}^{\perp\,\alpha}(z_1,{\bm b},\mu,\zeta)\tilde{H}_{1,h_2/{q}}^{\perp\,\beta}(z_2,{\bm b},\mu,\zeta)\bigg]\,.
\label{eq:diffcsb2}
\end{align}
To proceed, we will also have a ``subtracted'' TMD Collins FF $\tilde{H}_{1,h/{q}}^{\perp}(z,b,\mu,\zeta)$ following Eq.~\eqref{eq:Hperp(u)},
\begin{align}
\tilde{H}_{1,h_1/{q}}^{\perp\,\alpha}(z,{\bm b},\mu,\zeta) \equiv \left(-\frac{i{\bm b}^\alpha}{2}\right) \tilde{H}_{1,h/{q}}^{\perp}(z,b,\mu,\zeta)\,.
\end{align}
Next, we carry out the Lorentz contractions on $\alpha$ and $\beta$ and then perform the energy-fraction-weighted integrals in Eq.\ \eqref{eq:EECazdef2} to obtain the azimuthal-dependent EEC in the back-to-back limit and arrive at
\begin{align}
\mathrm{{{EEC}}}_{e^+e^-}(\tau,\phi)& = \frac{d\Sigma_{e^+e^-}}{d\tau d\phi}\nonumber\\
&=\frac{1}{2}\sigma_0\,H(Q,\mu) \sum_q e_q^2\int\frac{b\,db}{2\pi}
\bigg[J_0(b\sqrt{\tau} Q)\jet_{q}(b,\mu,\zeta)\jet_{\bar{q}}(b,\mu,\zeta)
\nnu
& \hspace{2.5cm}+ \cos (2{\phi}) \,\frac{b^2}{8}J_2(b \sqrt{\tau} Q)\left. \jet_{q}^{\perp}(b,\mu,\zeta)\jet_{\bar{q}}^{\perp}(b,\mu,\zeta)\right]\,.
\label{eq:epemEEC}
\end{align}
This is one of the key results of our paper, which shows that a new term $\propto \cos(2\phi)$ arises in the azimuthal dependent EEC observable for $e^+e^-$ annihiliation. Here it is important to emphasize that beside $\jet_q(b,\mu,\zeta)$, the so-called unpolarized EEC jet function introduced previously in~\cite{Moult:2018jzp}, we have defined a new EEC jet function $\jet_q^{\perp}(b,\mu,\zeta)$. The unpolarized EEC jet function has a close relation to the unpolarized TMD FFs, while $\jet_q^{\perp}(b,\mu,\zeta)$ is closely connected with the Collins fragmentation functions:
\begin{align}
\label{eq:unpjet}
\jet_q^{}(b,\mu,\zeta)\equiv&
\sum_{h}\int_0^1dz\,z\,\tilde{D}_{1,h/q}(z,b,\mu,\zeta)\,,
\\
\jet_q^{\perp}(b,\mu,\zeta)
\equiv&
\sum_{h}\int_0^1dz\,z\, \tilde{H}_{1,h/q}^{\perp}(z,b,\mu,\zeta)
\label{eq:Collinsjet}\,.
\end{align}
Because of this close connection, we name the new EEC jet function $\jet_q^{\perp}(b,\mu,\zeta)$ as the ``Collins-type'' EEC jet function. Note that the left-hand side in Eqs.~\eqref{eq:unpjet} and \eqref{eq:Collinsjet} no longer carries hadron label as we sum over the final state hadrons as appropriate from the inclusive nature of the EEC. We will defer the discussion on the properties of these jet functions, especially the Collins-type EEC jet function in Sec.~\ref{sec:jetfunc}. We will now study the azimuthal dependent EEC in $ep$ collisions.

\subsection{Azimuthal dependent EEC in DIS}
\label{sec:DISBreit}
\begin{figure}[h]
\centering
\includegraphics[width=0.49\textwidth,trim={0 0 5cm 0},clip]{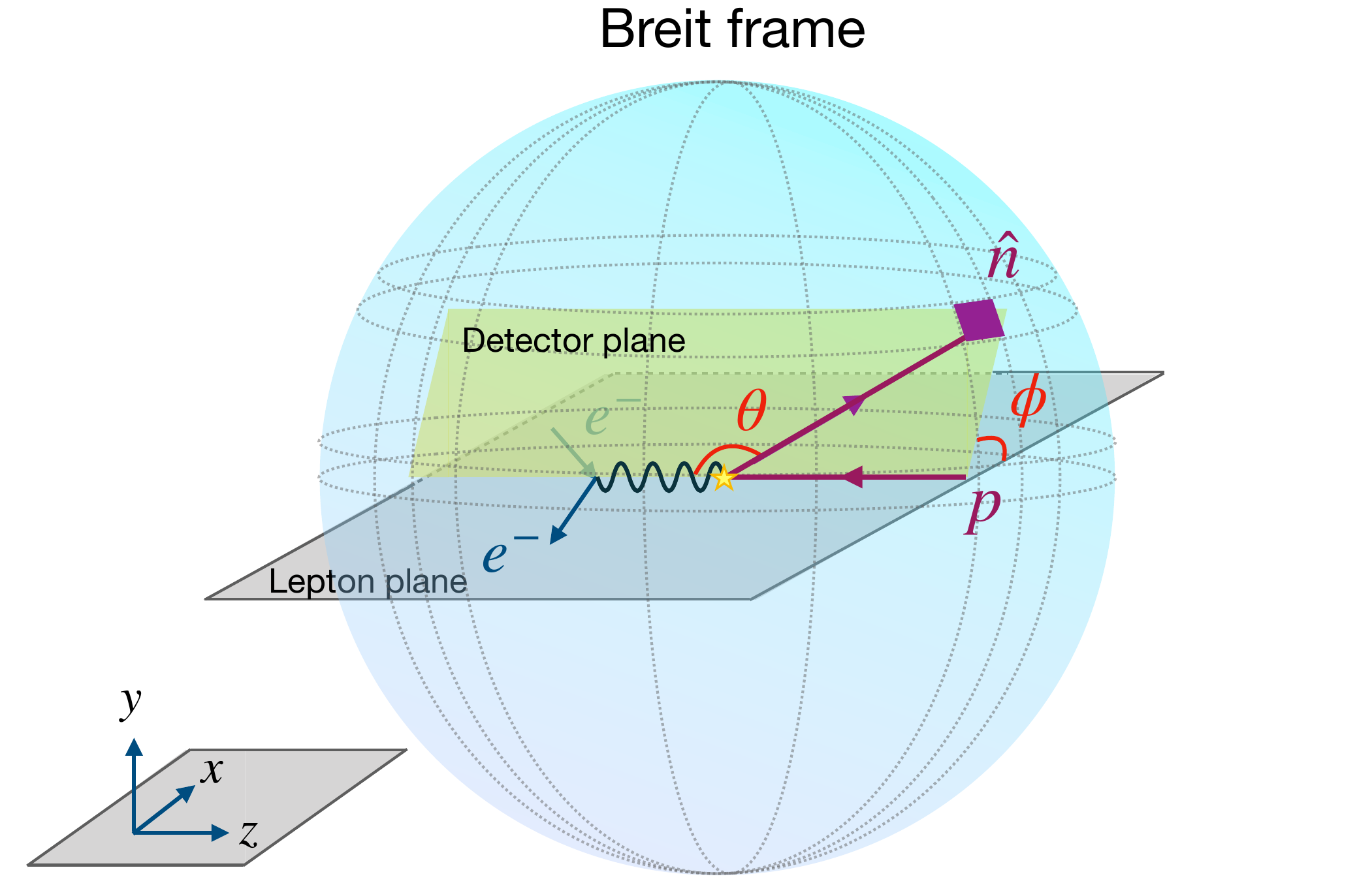}
\includegraphics[width=0.49\textwidth,trim={0 0 5cm 0},clip]{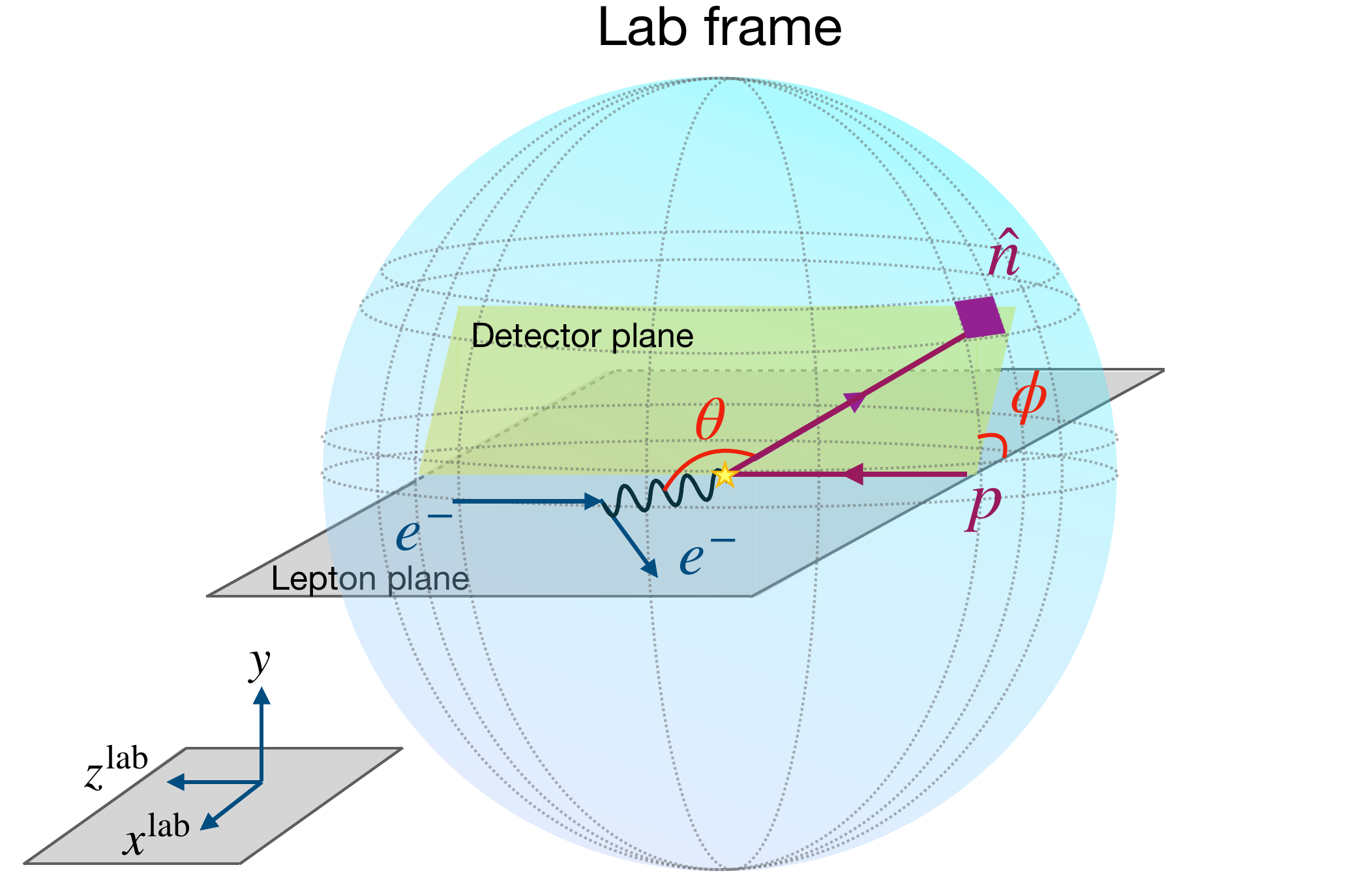}
\caption{\text{Left}: Illustration of {{EEC}} for DIS in the Breit frame using the Trento conventions. Here the exchanged virtual photon is along the $+z$ direction and the incoming proton moves along the $-z$ axis, together with the incoming lepton, they form the lepton plane (shown as gray in the figure). The detector direction $\hat{n}$ and the initial proton forms the detector plane (shown as yellow), which has an azimuthal angle $\phi$ with respect to the lepton plane. The angle between the detector and the proton $p$ is given by $\theta$. \text{Right}:  {{EEC}} for DIS in the EIC lab frame. Here the incoming proton is along the $+z^{\rm lab}$ direction and the incoming electron is along the $-z^{\rm lab}$ direction. }
\label{fig:eec_ep}
\end{figure}

A modified version of EEC in the deep inelastic $ep$ scattering was first introduced in~\cite{Li:2021txc} in the Breit frame. It measures the energy correlation as a function of the angle $\theta$ (we also often use $\tau$  defined similarly as Eq.~\eqref{eq:taudef} for convenience) between a detector and the incoming proton $p$ beam, as shown in the left panel of Fig.~\ref{fig:eec_ep}.\footnote{Technically, since we detect energy flowing into only a single detector, calling it an `energy-energy' correlator is somewhat of a misnomer.}  As in~\cite{Li:2021txc}, we begin by generalizing the DIS version of EEC in the Breit frame, which we define using the Trento conventions. In this frame, the exchanged virtual photon has no temporal component and points along the $+z$ direction and the proton beam moves along the $-z$ axis. We can then form two planes: the detector plane that is generated by the detector direction and the initial proton $p$ and the lepton plane that is formed by the incoming (outgoing) leptons and the initial proton. We denote the azimuthal angle difference between these two planes as $\phi$. We will be particularly interested in studying the consequence of such an azimuthal dependent EEC for DIS process, especially when both the incoming lepton and proton beams can be polarized. 

Let us start with the kinematics. Since the EEC will be closely related to the TMD factorization for producing a final-state hadron, $e(\ell)+p(P)\to e(\ell')+h(P_h)+X$, we define the usual semi-inclusive DIS kinematic variabbles, 
\begin{align}
    x_B = \frac{Q^2}{2P\cdot q},
    \qquad
    y = \frac{Q^2}{x_B S},
    \qquad
    z_h = \frac{P\cdot P_h}{P\cdot q},\label{eq:diskin}
\end{align}
where $q=\ell'-\ell$ is the momentum of the exchanged virtual photon with $Q^2 = -q^2$ and $S=(\ell+P)^2$ is the electron-proton center-of-mass energy. Denoting the corresponding opening and azimuthal angles of a given final-state hadron measured in the detector as $\theta_{hp}$ and $\phi_h$, we generalize the EEC definition in~\cite{Li:2021txc} to include the azimuthal angle dependence as
\begin{align}
\label{eq:EECazimDISdef}
\mathrm{{{EEC}}_{DIS}}(\tau,\phi) &\equiv \frac{d\Sigma_{\rm DIS}}{dx_B dy d\tau d\phi}\\
&\hspace{-1cm}=\sum_{h}\int d\theta_{hp}\,d\phi_{h} \,dz_h \,z_h \, \frac{d\sigma}{dx_B dy dz_h d\theta_{hp} d\phi_{h}}\, \delta\left(\tau - \frac{1+\cos \theta_{hp}}{2}\right)\delta(\phi - \phi_{h})\,,
\nonumber
\end{align}
where we have suppressed $x_B$ and $y$ dependence in the azimuthal dependent EEC definition, $\mathrm{{{EEC}}_{DIS}}(\tau,\phi)$. 

Similar to the discussion above for $e^+e^-$ collisions, in the back-to-back limit of the incoming proton and the detector, i.e. $\theta \to \pi$ (or $\tau \to 0$), we find
\begin{align}
\tau = \frac{\bm{P}_{hT}^2}{z_h^2 Q^2}\,,
\end{align}
for a given outgoing hadron inside the detector up to power corrections. Here $\bm{P}_{hT}$ is the transverse momentum of the final hadron $h$, measured using the Trento convention~\cite{Bacchetta:2004jz}  with azimuthal angle $\phi_h$. For convenience, we introduce the shorthand notation $\bm{q}_T \equiv -\bm{P}_{hT}/z_h$, which corresponds to the virtual photon's transverse momentum in the center-of-mass frame where the initial proton and final hadron align along the $z$-axis. From the Breit frame we use here, $\bm{q}_T$ should just be thought of as a convenient auxiliary variable. 

With the relation between $\tau$ and hadron's transverse momentum, we can then relate the azimuthal dependent EEC in Eq.~\eqref{eq:EECazimDISdef} in the back-to-back limit to the ${\bm q}_T$-differential cross section as
\begin{align}
\label{eq:EECazimDISdef2}
\mathrm{{{EEC}}}_{\rm DIS}(\tau,\phi) =&  \frac{d\Sigma_{\rm DIS}}{dx_B dy d\tau d\phi} 
\nnu
=&\sum_{h}\int d^2\bm{q}_T \,dz_h \,z_h\, \frac{d\sigma}{dx_B dy dz_h d^2\bm{q}_T } \,\delta\left(\tau - \frac{\bm{q}_T^2}{Q^2}\right) \delta(\phi - \phi_h)\,,
\end{align}
where $\phi_h$ is the azimuthal angle of $\bm{P}_{hT}$. The standard TMD factorization in the Breit frame tells us that the transverse momentum can be sourced from different contributions as~\cite{Collins:2011zzd,Bacchetta:2022awv}
\begin{align}
\bm{q}_T = -\frac{\bm{p}_{\perp}}{z_h} - \bm{k}_{\perp} - \bm{\lambda}_{\perp}\,,
\end{align}
where $\bm{p}_{\perp}$ is the hadron transverse momentum with respect to the fragmenting parton (quark or anti-quark, in this case), $\bm{k}_{\perp}$ is the transverse momentum of the parton inside the incoming proton, and $\bm{\lambda}_{\perp}$ is the soft-radiation describing the recoil and is measured with respect to the photon-beam axis. 

Then in the back-to-back limit, using this known TMD factorization in the Breit frame (e.g. see~\cite{Kotzinian:1994dv,Diehl:2005pc,Bacchetta:2006tn,Anselmino:2008sga,Kang:2015msa}), the azimuthal angle dependent {{EEC}} for DIS defined in the Eq.\ \eqref{eq:EECazimDISdef} can be presented in the back-to-back limit as
\begin{align}
\mathrm{{{EEC}}}_{\rm DIS}(\tau,\phi)  &= \frac{d\Sigma_{\rm DIS}}{dx_Bdyd\tau d\phi}=\sigma_0\int d^2{\bm q}_T\delta(\tau -\frac{{\bm q}_T^2}{Q^2})\delta(\phi - \phi_{h})\int\frac{db\ b}{2\pi}\bigg\{\mathcal{F}_{UU}\nnu
&+\cos(2\phi_{h})\frac{2(1-y)}{1+(1-y)^2}\mathcal{F}_{UU}^{\cos(2\phi_{h})}+S_{\parallel}\sin(2\phi_{h})\frac{2(1-y)}{1+(1-y)^2}\mathcal{F}_{UL}^{\sin(2\phi_{h})}\nnu
&+|{\bm S}_\perp|\bigg[\sin(\phi_{h}-\phi_s)\mathcal{F}_{UT}^{\sin(\phi_{h}-\phi_s)}+\sin(\phi_{h}+\phi_s)\frac{2(1-y)}{1+(1-y)^2}\mathcal{F}_{UT}^{\sin(\phi_{h}+\phi_s)}\nnu
&\hspace{1.4cm}+\sin(3\phi_{h}-\phi_s)\frac{2(1-y)}{1+(1-y)^2}\mathcal{F}_{UT}^{\sin(3\phi_{h}-\phi_s)}\bigg]\nnu
&+\lambda_e\bigg[S_{\parallel}\frac{y(2-y)}{1+(1-y)^2}\mathcal{F}_{LL}
+|{\bm S}_\perp|\cos(\phi_{h}-\phi_s)\mathcal{F}_{LT}^{\cos(\phi_{h}-\phi_s)}\bigg]\bigg\}\,,\label{eq:EECDIS3}
\end{align} 
where the indices $A$ and $B$ of the structure functions $\mathcal{F}_{AB}$ represent the polarization of the incoming electron and incoming proton, respectively. The born cross-section is given by 
\begin{align}
\sigma_0=\frac{2\pi\alpha_{\rm em}^2}{Q^2}\frac{1+(1-y)^2}{y}\,.
\end{align}
Also, $S_{\parallel}$ and $|{\bm S}_\perp|$ respectively denote the helicity and transverse spin of the incoming proton, whereas $\lambda_e$ denotes the helicity of the incoming electron. The angle $\phi_s$ is the azimuthal angle of the transverse spin of the beam.
The exact expression of each structure functions in terms of different nuclear TMDs and the EEC jet functions defined above in Eqs.\ \eqref{eq:unpjet} and\ \eqref{eq:Collinsjet} are given in App.~\ref{app1}. Unlike the {{EEC}} in $e^+e^-$ annihilation, the DIS version of the {{EEC}} has only one EEC jet function per term as it measures correlation between one outgoing hadron and the beam.

\subsection{Azimuthal dependent EEC in DIS using the lab-frame angles}
\label{sec:labEEC}
As demonstrated in~\cite{Gao:2022bzi}, a new angular observable $q_*$ defined in the Lab frame was proposed for precisely extracting the transverse momentum distribution of hadrons in DIS experiments, by taking advantage of the near-perfect resolution on the \emph{angles} of charged particle tracks as opposed to momentum. Although energy correlators are inherently angular observables, boosting to the Breit frame in Sec.~\ref{sec:DISBreit} requires precise determination of the virtual photon momentum, which compromises precision. To this end, we study how energy correlator in DIS can be formulated using the angle defined in the lab frame using the proposed $q_*$ observable.

\begin{figure}[h]
\centering
\includegraphics[width=0.4\textwidth]{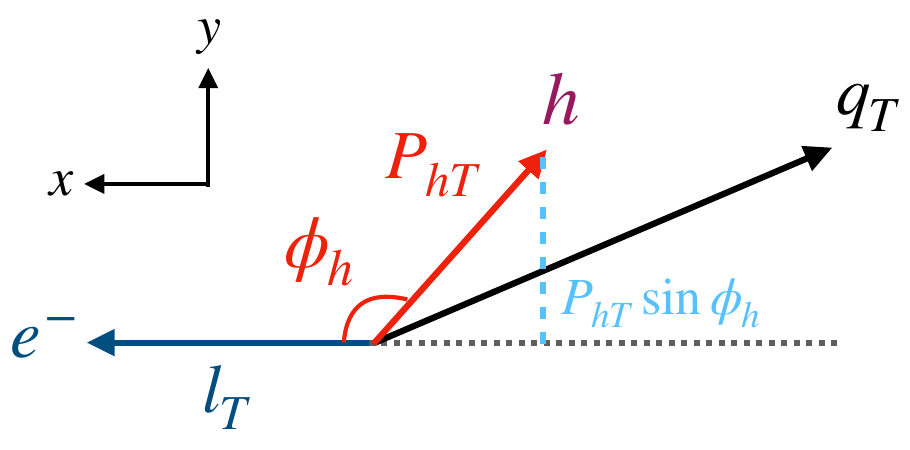}
\caption{Illustration of $\phi_h$ in the $x-y$ 
 plane of the left panel in Fig.~\ref{fig:eec_ep}.}\label{fig:eec_ep2}
\end{figure}

More specifically, as discussed in \cite{Gao:2022bzi}, a high-precision reconstruction of polar and azimuthal angles in the EIC lab frame was developed to exploit the acoplanarity angle $\phi_{\rm acop}^{\mathrm{EIC}}$ as a precision probe of hadron transverse momentum $\bm{P}_{hT}$ measured in Trento convention discussed above. In the EIC lab frame as given in right panel of Fig.~\ref{fig:eec_ep}, the acoplanarity angle $\phi_{\rm acop}^{\mathrm{EIC}}$ are defined as $P_{hy}^{\rm EIC}/P_{hx}^{\rm EIC}$, where $P_{hx}^{\rm EIC}$ and $P_{hy}^{\rm EIC}$ is the $x$ and $y$ component of the produced hadron momentum in the EIC frame. At the leading power kinematics, with $\sqrt{\bm{P}_{hT}^2} = P_{hT}$, one can relate this angle with quantities in the Breit frame as
\begin{align}
    \tan \phi_{\rm acop}^{\mathrm{EIC}} =-\frac{P_{hT}\sin\phi_h}{z_hQ\sqrt{1-y}}+\mathcal{O}\left({P_{hT}^2}/{(z_hQ)}^2\right) = -\sqrt{\frac{\tau}{1-y}}\sin\phi_h + \mathcal{O}(\tau)\,,\label{eq:tanphiacop}
\end{align}
where $Q^2=-q^2=-(l-l')^2$, and $y$ and $z_h$ have been defined in Eq.~\eqref{eq:diskin}. To probe $P_{hT}$ more straightforwardly with quantities that can be directly measured in the EIC frame, an optimized observable $q_*$ is constructed and given by
\begin{align}
q_* \equiv 2 P_{\mathrm{EIC}}^0 \frac{e^{\eta_h}}{1+e^{\Delta \eta}} \tan \phi_{\mathrm{acop}}^{\mathrm{EIC}}    \,.
\end{align}
 Here $P_{\rm EIC}^0$, $\eta_h$ and $\eta_e$ are energy and pseudorapidities measured in the EIC frame. At the leading-power limit $P_{hT} \ll z_h Q$, one can apply the relation between $\phi_{\rm acop}^{\mathrm{EIC}}$ and $P_{hT}$ in Eq.~\eqref{eq:tanphiacop} and obtain the approximation $q_*\approx -P_{hT}\sin\phi_h$ to simplify the analytical calculation. An illustration of $\phi_h$ in the $x$-$y$ plane of Fig.~\ref{fig:eec_ep} is shown in Fig.~\ref{fig:eec_ep2}. Although $\phi_h$ here, the azimuthal angle of ${\bm P}_{hT}$, is measured in the Trento frame, the observable $q_*$ itself is purely a measurable observable in the lab frame and designed to be maximally resilient against resolution effects while delivering the same sensitivity to TMD dynamics as $P_{hT}\ll Q$. Moreover, this resolves the issue of accurately reconstructing small transverse momentum $P_{hT}$ in 3d measurements of confinement and hadronization. 

Now we will construct a new EEC observable called $\mathrm{{{EEC}}}_{\rm DIS}^*$ as a function of $q_*$,
\begin{align}
\mathrm{{{EEC}}}_{\rm DIS}^*(q_*)=\frac{d\Sigma_{\rm DIS}}{dx_Bdydq_*}=& \sum_{h} \int dz_h z_h \frac{d\sigma}{dx_Bdydz_hdq^h_*} \delta(q_* - q_*^h)\,.
\end{align}
Using the factorization theorem presented for $\frac{d\sigma}{dxdydz_hdq^h_*}$ given in~\cite{Gao:2022bzi}, the $\mathrm{{{EEC}}}_{\rm DIS}^*(q_*)$ is given as
\begin{align}
\frac{d\Sigma_{\rm DIS}}{dx_Bdydq_*}=&\sigma_{0}\int db\bigg\{\cos\left(bq_*\right)\bigg[\mathcal{C}[\tilde{f}_1\jet]+\frac{2(1-y)}{1+(1-y)^2}\mathcal{C}^{\perp}[\tilde{h}_1^{\perp(1)}\jet^{\perp}]+\lambda_eS_{\parallel}\frac{y(2-y)}{1+(1-y)^2}\mathcal{C}[\tilde{g}_{1L}\jet]\bigg]\nnu
&+|{\bm S}_\perp|\sin\left(bq_*\right)\cos\phi_s\bigg[-\mathcal{C}[\tilde{f}^{\perp(1)}_{1T}\jet]+\frac{2(1-y)}{1+(1-y)^2}\mathcal{C}^{\perp}[\tilde{h}_1\jet^{\perp}]\nnu
&+\frac{2(1-y)}{1+(1-y)^2}\frac{1}{4}\mathcal{C}^{\perp}[\tilde{h}^{\perp(2)}_{1T}\jet^{\perp}]\bigg]+\lambda_e|{\bm S}_\perp|\sin\left(bq_*\right)\sin\phi_s\mathcal{C}[\tilde{g}^{(1)}_{1T}\jet]\bigg\}\,,\label{eq:EECDISq20}
\end{align}
where the $\tilde{f}^{(m)}$ is $m$-moment TMD PDFs in the Fourier $b$-space ``subtracted'' with squared root of the soft function $\sqrt{S({\bm b}^2,\mu,\nu)}$, similar to the ``subtracted'' TMD FFs defined in Eq.~\eqref{eq:subTMD FFs}. $\jet$ and $\jet^{\perp}$ are the ``subtracted'' unpolarized and Collins-type EEC jet function, respectively, given in Eq.~\eqref{eq:unpjet} and \eqref{eq:Collinsjet}. For convenience, we define the notations $\mathcal{C}^{(\perp)}[\tilde{f}^{(m)}\jet^{(\perp)}]$ as
\begin{align}
\mathcal{C}[\tilde{f}^{(m)}\jet]=&\sum_q e_q^2H(Q,\mu)\jet_q(b,\mu,\zeta) b^{m}{M^m}\tilde{f}^{(m)}_q(x,b,\mu,\zeta)\,,\label{eq:djq}\\
\mathcal{C}^{\perp}[\tilde{f}^{(m)}\jet^{\perp}]=&\sum_q e_q^2H(Q,\mu)\jet_q^\perp(b,\mu,\zeta)\frac{b^{m+1}}{2}{M^m}{(-M_h)}\tilde{f}^{(m)}_q(x,b,\mu,\zeta) \,.\label{eq:dpjq}
\end{align}
Using the new {{EEC}} observable defined in Eq.~\eqref{eq:EECDISq20}, one can get access to probing 7 leading-twist TMD PDFs with $q_*$ measured, which is constructed as a high-precision probe in the EIC lab frame. Note that the reason $\tilde{h}_{1L}^{\perp(1)}$ is not present in Eq.~\eqref{eq:EECDISq20} is that its coefficient $\sin(2\phi_h)$ has an odd symmetry thus the contribution vanishes after integrating over $\phi_h$.

\section{Properties of Collins-type EEC jet functions}\label{sec:jetfunc}
In this section, we derive the intricate behavior of the EEC jet functions using the operator product expansion (OPE) as done for the unpolarized case in~\cite{Moult:2018jzp}, and extend their analysis to the Collins-type EEC jet function defined in Eq.~\eqref{eq:Collinsjet}. We find that the Collins-type EEC jet function becomes null in the OPE region upon neglecting the off-diagonal matching terms, which is $\alpha_s$ suppressed relative to the diagonal terms. Finally, we conclude by discussing the application of considering less inclusive EEC jet function by restricting to a subset of hadrons. 

\subsection{Operator product expansion region}
In order to elucidate the OPE of the EEC jet functions, we start with the concept of ``subtracted" TMD FFs as defined in Eq.~\eqref{eq:subTMD FFs}, which absorbs the square-root of the soft-function. This serves as a vital starting point for our analysis, enabling us to uncover the intricate interplay between the OPE and the azimuthal angle dependence of the EEC jet functions.

The OPE of the subtracted unpolarized and Collins TMD FFs then gives\ \cite{Koike:2006fn,Yuan:2009dw,Bacchetta:2013pqa,Echevarria:2014rua,Kang:2015msa,Boussarie:2023izj}
\begin{align}
\label{eq:OPEunp}
\tilde{D}_{1,h/q}(z,b,\mu,\zeta)=&\left[C_{j\leftarrow q}\otimes D_{1,h/{j}} \right]\left(z,b,\mu,\zeta\right) + \mathcal{O}(b^2\Lambda_{\rm QCD}^2),\\
\tilde{H}_{1,h/{q}}^{\perp}(z,b,\mu,\zeta)=&\left[\delta C_{j\leftarrow q}^{\rm Collins}\otimes \hat{H}_{1,h/j}^{\perp(3)} + A_{j\leftarrow q}\tilde{\otimes} \hat{H}_{F,h/j}\right]\left(z,b,\mu,\zeta\right)+ \mathcal{O}(b^2\Lambda_{\rm QCD}^2)\,,
\label{eq:OPECollins}
\end{align}
where the usual convolution $\otimes$ is given by
\begin{align}
\left[C_{j\leftarrow {q}}\otimes F_{h/j}\right]\left(z,b,\mu,\zeta\right) = \int_{z}^{1} \frac{dx}{x} C_{j\leftarrow {q}}\left(\frac{z}{x}, b,\mu,\zeta \right) F_{h/j}\left(x, \mu\right)\,,\label{eq:defconvo}
\end{align}
and the convolution $\tilde{\otimes}$ is a double convolution 
\begin{align}
&\left[A_{j\leftarrow q}\tilde{\otimes} \hat{H}_{F,h/j}\right]\left(z,b,\mu,\zeta\right) \nnu
&\quad= \int_{z}^{1}\frac{dx}{x}\, \int\frac{dz_1}{z_1^2}\,{\rm PV}\left(\frac{1}{\frac{1}{x}-\frac{1}{z_1}}\right) A_{j\leftarrow {q}}\left(\frac{z}{x}, z_1,b,\mu,\zeta \right) \hat{H}_{F,h/j}\left(x,z_1, \mu\right)\,,\label{eq:a_hf}
\end{align}
where $A_{q'\leftarrow q}\left(z, z_1,\mu, \zeta \right)$ starts at the order $\mathcal{O}(\alpha_s)$ and will be ignored in our analysis relative to the $\delta C_{j\leftarrow q}^{\rm Collins}$ term. Also, ${\rm PV}$ represents principle value and Eq.~\eqref{eq:a_hf} comes from the two-variable twist-3 fragmentation function $\hat{H}_{F,h/q}(z_1,z_2,\mu)$ that appears in the collinear factorization region in $b \ll 1/\Lambda_{\rm QCD}$~\cite{Yuan:2009dw}. Such two-variable dependence arise from different longitudinal momentum flow on the amplitude and conjugate amplitude side.

Here, $D_{1,h/j}(z,\mu)$ in Eq.\ \eqref{eq:OPEunp} describes the unpolarized collinear fragmentation functions. On the other hand, $\hat{H}_{1,h/j}^{\perp (3)}(z,\mu)$ is a twist-3 fragmentation function, which can be related to the first ${\bm k}_\perp$-moment of the Collins TMD FF through an equation of motion relation~\cite{Yuan:2009dw,Kang:2015msa}.
Matching of the unpolarized TMD FF and the Collins TMD FF given in Eqs.\ \eqref{eq:OPEunp} and \eqref{eq:OPECollins} can be derived through the usual OPE in the small-$b$ region, $1/b \gg \Lambda_{\rm QCD}$. To one-loop, these matching coefficients of Eqs.\ \eqref{eq:OPEunp} and\ \eqref{eq:OPECollins} are given by (see e.g.~\cite{Bacchetta:2013pqa,Echevarria:2014rua,Kang:2015msa,Echevarria:2016scs}) 
\begin{align}
C_{q^{\prime} \leftarrow q}\left(z, b,\mu,\zeta\right)=&\delta_{q q^{\prime}}\bigg\{\delta\left(1-z\right)+\frac{\alpha_{s}}{\pi}\bigg[C_{F}\delta\left(1-z\right)\left(-\frac{L_b^2}{4}+\frac{L_b}{2}\left(\frac{3}{2}+\ln\frac{\mu^2}{\zeta^2}\right)-\frac{\pi^2}{24}\right)\nnu
&\qquad+\frac{C_{F}}{2}\left(1-z\right)+ \left(\ln z-\frac{L_b}{2}\right)P_{q \leftarrow q}\left(z\right)\bigg]\bigg\}+\mathcal{O}(\alpha_s^2)\,,\\
\delta C_{q'\leftarrow q}^{\rm Collins}\left(z, b,\mu,\zeta\right)=&\delta_{q q^{\prime}}\bigg\{\delta\left(1-z\right)+\frac{\alpha_{s}}{\pi}\bigg[C_{F}\delta\left(1-z\right)\left(-\frac{L_b^2}{4}+\frac{L_b}{2}\left(\frac{3}{2}+\ln\frac{\mu^2}{\zeta^2}\right)-\frac{\pi^2}{24}\right)\nnu
&\qquad+ \left(\ln z-\frac{L_b}{2}\right)\hat{P}^c_{q \leftarrow q}\left(z\right)\bigg]\bigg\}+\mathcal{O}(\alpha_s^2)\,,\label{eq:collins_c}
\end{align}
where $L_b=\ln(\mu^2/\mu_b^2)$ with $\mu_b = 2e^{-\gamma_E}/b$. The splitting functions $P_{q \leftarrow q}(z)$ and $\hat{P}_{q \leftarrow q}^{c}(z)$ is given as
\begin{align}
P_{q \leftarrow q}(z)&=C_{F}\left[\frac{1+z^{2}}{(1-z)_{+}}+\frac{3}{2} \delta(1-z)\right]\,,\\
\hat{P}_{q \leftarrow q}^{c}(z)&=C_{F}\left[\frac{2 z}{(1-z)_{+}}+\frac{3}{2} \delta(1-z)\right]\,.\label{eq:collins_pqq}
\end{align}
The matching coefficients $\delta C_{q'\leftarrow q}^{\rm Collins}$ from (anti-)quark to gluon or different flavored (anti-)quark are zero at $1$-loop for the Collins TMD FFs. And other matching coefficients for the unpolarized TMD FFs have been provided in~\cite{Kang:2015msa}. 

The collinear functions in the matching obey the sum rules \cite{Schafer:1999kn,Meissner:2010cc,Metz:2016swz}
\begin{align}
\sum_{h} \int_0^1 dz\,z\,D_{1,h/j}\left(z, \mu\right) = 1\,,\label{eq:unpsum}\\
\sum_{h}\int_0^1 dz\, \hat{H}_{1,h/q}^{\perp(3)}\left(z,\mu\right) = 0\,.
\label{eq:STsum}
\end{align}
The first sum rule is just the longitudinal momentum conservation, i.e. sum over longitudinal momentum fraction carried by the hadron $h$ is $1$. The second sum rule for the twist-$3$ fragmentation function is called Sch\"{a}fer-Teryaev (ST) sum rule, which is related to the transverse momentum conservation of the Collins TMD FF\ \cite{Schafer:1999kn}. A general proof of the ST sum rule in QCD was given in~\cite{Meissner:2010cc}. It was intuitively understood as the fact that the transverse momentum carried by the final hadron must sum to $0$ as the fragmenting parton has a $0$ transverse momentum. 

With the ``subtracted'' EEC jet function defined in Eqs.~\eqref{eq:unpjet} and~\eqref{eq:Collinsjet},
we provide the parameterization of our TMD FFs using the $b^*$ prescription~\cite{Collins:2011zzd,Aybat:2011zv,Prokudin:2015ysa,Kang:2015msa}, which ensures $\mu_{b_*} \gg \Lambda_{\rm QCD}$ and gives 
\begin{align}
\label{eq:Dsub}
\tilde{D}_{1,h/q}(z,b,\mu,\zeta_f)=&\,\tilde{D}_{1,h/q}(z,b,\mu_{b_*},\zeta_i) e^{-S_{\rm pert}(\mu,\mu_{b_*}) - S_{\rm NP}^{D_1}(b,Q_0,\zeta_f)}\left(\sqrt{\frac{\zeta_f}{\zeta_i}}\right)^{\kappa\left(b, \mu_{b_*}\right)}\,,\\
\tilde{H}_{1,h/{q}}^{\perp}(z, b,\mu,\zeta_f)=&\,
\tilde{H}_{1,h/{q}}^{\perp}(z,b,\mu_{b_*},\zeta_i)e^{-S_{\rm pert}(\mu,\mu_{b_*}) - S_{\rm NP}^{H^{\perp}_{1}}(b,Q_0,\zeta_f)}\left(\sqrt{\frac{\zeta_f}{\zeta_i}}\right)^{\kappa\left(b, \mu_{b_*}\right)}\,,
\label{eq:Hsub}
\end{align}
where $\kappa\left(b, \mu_{b_*}\right)$ is the Collins-Soper (CS) kernel~\cite{Collins:1981va}, denoting the universal rapidity evolution~\cite{Chiu:2011qc} for TMDs, and can be  perturbatively calculated at small $b$ region~\cite{Echevarria:2016scs}. When $b$ is large, the CS kernel is non-perturbative and can be computed in lattice QCD~\cite{Avkhadiev:2023poz,LatticePartonLPC:2023pdv,Shu:2023cot,Schlemmer:2021aij} or extracted from experimental data~\cite{Moos:2023yfa,Bacchetta:2022awv,Bertone:2019nxa,Bacchetta:2019sam,Scimemi:2019cmh}.
The non-perturbative Sudakov factors $S_{\rm NP}^{D_1}$ and $S_{\rm NP}^{H^{\perp}_{1}}$ in general depend on the hadron species and we will explore it in the section below. 

In\ \cite{Moult:2018jzp}, interesting observation was made about simplification of unpolarized EEC jet function when the non-perturbative Sudakov effects are ignored, i.e. $\frac{1}{b} \gg \Lambda_{\rm QCD}$. That is, using the sum rule in Eq.\ \eqref{eq:unpsum}, we arrive at
\begin{align}
\jet_q(b,\mu,\zeta)=&\sum_{h}\int_0^1 z\,dz\, \tilde{D}_{1,h/q}(z,b,\mu,\zeta)\nnu
=& \sum_{h}\int_0^1 z\,{dz}  \int_{z}^1 \frac{dx}{x}\, C_{j\leftarrow q}\left(\frac{z}{x}, b,\mu_{b_*},\zeta \right) D_{1,h/j}\left(x, \mu_{b_*}\right) e^{-S_{\rm pert}(\mu,\mu_{b_*})} \nnu
=& \int_0^1 \tau\,{d\tau}\,{C}_{j\leftarrow q}(\tau,b,\mu_{b_*},\zeta)\left[\sum_{h} \int_0^1 {dx}\,x\,D_{1,h/j}\left(x, \mu_{b_*}\right)\right]e^{-S_{\rm pert}(\mu,\mu_{b_*})}   \nnu
=& \int_0^1 \tau\,{d\tau}\,{C}_{j\leftarrow q}(\tau,b,\mu_{b_*},\zeta)e^{-S_{\rm pert}(\mu,\mu_{b_*})} \,, \label{eq:sumunp}
\end{align}
and thus reproduce the result from\ \cite{Moult:2018jzp} that unpolarized EEC jet function is purely perturbative object in the OPE region, given by the matching coefficients $C_{j\leftarrow q}$ alone.

On the other hand, the Collins-type EEC jet function can be simplified analogously using the ST sum rule described in Eq.~\eqref{eq:STsum}, with the contribution from twist-3 function $\hat{H}_{F,h/j}$ in Eq.~\eqref{eq:OPECollins} neglected. By making this simplification, we can achieve the same level of accuracy as in the previous scenario. In the OPE region, the Collins-type EEC jet functions can be further simplified as follows:
\begin{align}
\jet_q^{\perp}(b,\mu,\zeta) = &\,\sum_{h}\int_0^1dz\,z\, \tilde{H}_{1,h/q}^{\perp}(z,b,\mu,\zeta)
\nnu
=&\,\sum_{h}\int_0^1dz\int_{z}^1 \frac{dx}{x}\, \delta C_{q\leftarrow q}^{\rm Collins}(\frac{z}{x},b,\mu_{b_*},\zeta) \hat{H}_{1,h/q}^{\perp(3)}\left(x,\mu_{b_*}\right)e^{-S_{\rm pert}(\mu,\mu_{b_*})}\nnu
=&\,\int_0^1 d\tau \delta C_{q\leftarrow q}^{\rm Collins}(\tau,b,\mu_{b_*},\zeta) \left[\sum_{h}\int_0^1 dx\,\hat{H}_{1,h/q}^{\perp(3)}\left(x,\mu_{b_*}\right)\right]e^{-S_{\rm pert}(\mu,\mu_{b_*})}\nnu
=&\,\,0\,,
\label{eq:sumperp}
\end{align}
which demonstrates that the Collins-like EEC jet function $\jet_q^\perp$ becomes $0$ in the OPE region. Both of these results only hold true when non-perturative effects can be ignored. 


\subsection{Collins-type EEC with subsets of hadrons}
\label{subsec:EECsubset}
In this section, we explore a less inclusive version of EEC in the back-to-back limit that is only sensitive to the energy flow of subset of hadrons $\langle \mathcal{E}_{\mathbb{S}_1}(\hat{n}_1)\mathcal{E}_{\mathbb{S}_2}(\hat{n}_2)\rangle$.\footnote{See~\cite{Lee:2023tkr,Lee:2023npz,Li:2021zcf,Chen:2020vvp,Jaarsma:2023ell} for similar consideration in the context of EEC in the collinear limit, where the track function formalism was used to study energy correlation between hadrons with specific quantum number.} Here, we denote some subset of hadron sharing some quantum number as $\mathbb{S}$. By considering the $\tau,\ \phi$-differential and $z$-weighed cross section, while summing over this subset of hadrons $\mathbb{S}$, we arrive at the modified jet functions from Eqs.\ \eqref{eq:unpjet} and\ \eqref{eq:Collinsjet},
\begin{align}
\label{eq:subJ}
\jet_{q/\mathbb{S}}(b,\mu,\zeta)\equiv&\sum_{h \in \mathbb{S}}\int_0^1dz \,z\,\tilde{D}_{1,h/q}(z,b,\mu,\zeta)\,,\\
\jet_{q/\mathbb{S}}^{\perp}(b,\mu,\zeta)\equiv&\sum_{h \in \mathbb{S}}\int_0^1dz \,z\, \tilde{H}_{1,h/q}^{\perp}(z,b,\mu,\zeta)\,.
\label{eq:subJperp}
\end{align}

In the OPE region, these subset jet functions can then be matched to the collinear functions as
\begin{align}
\jet_{q/\mathbb{S}}(b,\mu,\zeta)=& \sum_j F_{j\to \mathbb{S}} \int_0^1  {d\tau}\, \tau\,{C}_{j\leftarrow q}(\tau,b,\mu_{b_*},\zeta)e^{-S_{\rm pert}(\mu,\mu_{b_*})}\,,\label{eq:subJOPE}\\
\jet_{q/\mathbb{S}}^{\perp}(b,\mu,\zeta)=&\sum_j F^{\perp}_{j\to \mathbb{S}} \int_0^1 d\tau \,\delta C_{j\leftarrow q}^{\rm Collins}(\tau,b,\mu_{b_*},\zeta)e^{-S_{\rm pert}(\mu,\mu_{b_*})}\,,\label{eq:subJperpOPE}
\end{align}
Physically, $F_{j\to \mathbb{S}}$ is the average fraction of longitudinal momentum of parton $j$ carried by the subset $\mathbb{S}$ of hadrons~\cite{Li:2021txc}, which is thus between $0$ and $1$. On the other hand, $F^{\perp}_{j\to \mathbb{S}}$ is related to the average transverse momentum carried by the subset $\mathbb{S}$ of the hadrons, which is thus able to take any real value. These two functions are defined in terms of the moments of NP collinear objects as
\begin{align}
F_{j\to \mathbb{S}}&=\sum_{h\in \mathbb{S}} \int_0^1 {dz}\,z\,D_{1,h/j}\left(z, \mu_{b_*}\right)\,,\\
F^{\perp}_{j\to \mathbb{S}}&=\sum_{h\in\mathbb{S}}\int_0^1 dz\,\hat{H}_{1,h/j}^{\perp(3)}\left(z,\mu_{b_*}\right)\,.
\end{align}
It is worth noting that since $\delta C_{j\leftarrow q}^{\rm Collins}$ is only non-zero when $j=q$, the only relevant term is $F^{\perp}_{q\to \mathbb{S}}$. Additionally, as demonstrated in Eqs.~\eqref{eq:unpsum} and~\eqref{eq:STsum}, when $\mathbb{S} =$ {\it{all hadrons}}, one will obtain $F_{j\to \mathbb{S}}=1$ and $F^{\perp}_{j\to \mathbb{S}}=0$. 

Although the EEC jet functions are now sensitive to the non-perturbative information carried by the subset of hadrons, it is worth noting that all of the non-perturbative information are captured by a single number in the OPE region. As pointed out in~\cite{Li:2021txc}, considering {{EEC}} with subsets of hadrons can be interesting from the point of view of considering a set of all charged particles $\mathbb{S} = \mathbb{C}$ or a set of a single hadron type $\mathbb{S} = h$. From the perspective of Collins-type EEC jet function, we are motivated to consider the so-called {\it favored} and {\it unfavored} subset as well\ \cite{Vogelsang:2005cs,Kang:2015msa,Metz:2016swz}. The vanishing value of Collins-type EEC jet function in the OPE region can be understood as $F^{\perp}_{j\to \text{fav}} \approx - F^{\perp}_{j\to  \text{unfav}}$, and thus it is also interesting to consider $\mathbb{S} = \{\pi^+\}, \{\pi^-\},\{\pi^0\}, \{\pi^+,\pi^-\},\{\pi^+,\pi^-,\pi^0\}$ for Collins-type EEC in phenomenology.

\section{Phenomenology}\label{sec:pheno}
The present study focuses on the Collins azimuthal asymmetry in $e^+e^-$ annihilation with subsets of hadrons, namely $\mathbb{S}=\{\pi^+,\ \pi^-\}$, $\{\pi^+\}$, and $\{\pi^-\}$, as the hadron component of the EEC jet using EIC kinematics. Furthermore, we provide predictions for both the Collins and Sivers asymmetry in the Breit frame for the DIS process. It is worth noting that the EEC jet function is also a powerful tool for probing the internal structure of nucleons in the lab frame, as evidenced by our example of an azimuthal asymmetry related to the worm-gear TMD PDFs and EEC jet functions.

\subsection{The EEC Collins asymmetry in $e^+e^-$ annihilation}\label{sec:phenoee}
With the EEC derived for $e^+e^-$ annihilation in the back-to-back region in terms of the EEC jet functions in Eq.\ \eqref{eq:epemEEC}, we now carry out phenomenological study of the asymmetry associated with {{EEC}}. To achieve this, we first rewrite the Eq.\ \eqref{eq:epemEEC} as
\begin{align}
\mathrm{{{EEC}}}_{e^+e^-}(\tau,\phi)=\frac{d\Sigma_{e^+e^-}}{d\tau d\phi}=&\frac{1}{2}\sigma_0\sum_q e_q^2\int d{\bm q}_T^2\,\delta(\tau  - \frac{{\bm q}_T^2}{Q^2})\, Z_{uu}\,\left[1+ \cos(2\phi)\, \frac{Z_{\rm Collins}}{Z_{uu}}\right]\nonumber\\
\equiv&\frac{1}{2}\sigma_0\sum_q e_q^2\, Z_{uu}\,\left[1+ \cos(2\phi)\, A_{e^+e^-}(\tau Q^2)\right]\,,
\end{align}
where
\begin{align}
Z_{uu}=&\int\frac{bdb}{2\pi}J_0(bq_T)\jet_{q}(b,\mu,\zeta)\jet_{\bar{q}}(b,\mu,\zeta)\,,\label{zuu}\\
Z_{\rm Collins}=&\int\frac{bdb}{2\pi}\frac{b^2}{8}J_2(bq_T) \jet_{q}^{\perp}(b,\mu,\zeta)\jet_{\bar{q}}^{\perp}(b,\mu,\zeta)\,.\label{zcol}
\end{align}
Recall that subtracted EEC jet functions were defined in Eqs.\ \eqref{eq:unpjet} and\ \eqref{eq:Collinsjet}. As pions are the primary measurements available for the Collins-type asymmetry, we now study the ratio $A_{e^+e^-}^{\mathbb{S}_1\times\mathbb{S}_2}(\bm{q}_T^2) =A_{e^+e^-}^{\mathbb{S}_1\times\mathbb{S}_2}(\tau Q^2) $ with different subsets $\mathbb{S}_1\times\mathbb{S}_2$ of produced pion pairs. This amounts to simply using EEC jet functions with a subset in Eqs.\ \eqref{eq:subJ} and\ \eqref{eq:subJperp} defined in terms of TMD FFs of the pions.
As explained in the section\ \ref{sec:phenoee}, TMD FFs can be matched to the collinear functions in the OPE region. 
We now use the parametrization of \cite{Kang:2015msa} to generate $A_{e^+e^-}^{\mathbb{S}_1\times\mathbb{S}_2}(\tau Q^2)$. For the TMD Collins FFs in Eq.~\eqref{eq:Hsub}, at next-to-leading logarithm (NLL), we have 
\begin{align}
    \tilde{H}_{1,h/{q}}^{\perp}(z,b,\mu,\zeta)=&\left[\delta C_{j\leftarrow q}^{\rm Collins}\otimes \hat{H}_{1,h/j}^{\perp(3)}\right]\left(x,b,\mu_{b_*},\mu_{b_*}^2\right)e^{-\frac{1}{2}S_{\rm pert}\left(\mu, \mu_{b_*} \right) - S_{\rm NP}^{H_1^\perp}\left(b, Q_0, \zeta\right)}\,.
\end{align}
The corresponding twist-3 collinear fragmentation functions $\hat{H}^{(3)}(z,Q_0)$ were parametrized as
\begin{align}
\hat{H}_{fav}^{(3)}(z,Q_0)&=N_u^cz^{\alpha_u}(\tau )^{\beta_u}D_{1,\pi^+/u}(z,Q_0)\,,\label{eq:kang15_1}\\
\hat{H}_{unfav}^{(3)}(z,Q_0)&=N_d^cz^{\alpha_d}(\tau )^{\beta_d}D_{1,\pi^+/d}(z,Q_0)\,,\label{eq:kang15_2}\\
\hat{H}_{s/\bar{s}}^{(3)}(z,Q_0)&=N_d^cz^{\alpha_d}(\tau )^{\beta_d}D_{1,\pi^+/s,\bar{s}}(z,Q_0)\,,\label{eq:kang15_3}
\end{align}
and fitting parameters $N_{u}^c,\ N_{d}^c,\ \alpha_u,\ \alpha_d,\ \beta_u,\ \beta_d,$ are provided in \cite{Kang:2015msa}. The collinear FFs $D_{1,h/i}$ are the NLO $\mathtt{DSS}$ fragmentation functions~\cite{deFlorian:2014xna}. 
Note here the contribution from twist-3 function $\hat{H}_{F,h/j}$ in Eq.~\eqref{eq:OPECollins} has been neglected following what was done in~\cite{Kang:2015msa} and the coefficient $\delta \hat{C}^{\rm Collins}_{q^{\prime} \leftarrow q}$ is given in Eq.~\eqref{eq:collins_c}. 
Here we also provide the parametrization of the non-perturbative Sudakov factors for unpolarized and Collins TMD FFs,
\begin{align}
S_{\rm NP}^{D_1}(b,Q_0,\zeta)&=\frac{g_2}{2}\ln\left(\frac{b}{b_*}\right)\ln\left(\frac{\sqrt\zeta}{Q_0}\right)+\frac{g_h}{z^2} b^2\,,\label{eq:snpd1}\\
S_{\rm NP}^{H_1^\perp}(b,Q_0,\zeta)&=\frac{g_2}{2}\ln\left(\frac{b}{b_*}\right)\ln\left(\frac{\sqrt\zeta}{Q_0}\right)+\frac{g_h-g_c}{z^2} b^2\,,\label{eq:snph1p}
\end{align}
where $g_2=0.84$, $Q_0^2=2.4$ GeV$^2$, $g_h = 0.042$ GeV$^2$ and $g_c$ is extracted in~\cite{Kang:2015msa}. 
\begin{figure}[h]
\centering
\includegraphics[width=0.54\textwidth]{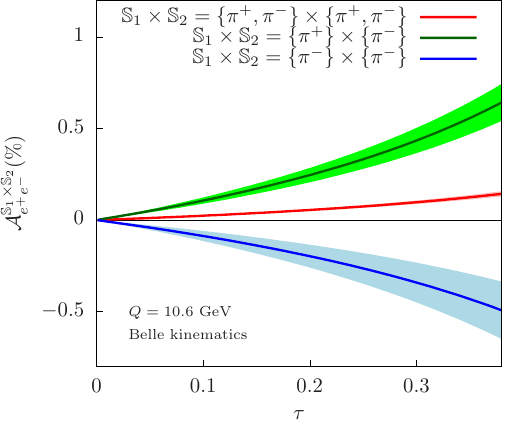}
\caption{$A_{e^+e^-}^{\mathbb{S}_1\times\mathbb{S}_2}$ for $\mathbb{S}_1\times\mathbb{S}_2=\{\pi^+,\ \pi^-\}\times\{\pi^+,\ \pi^-\}$, $\{\pi^+\}\times\{\pi^-\}$ and $\{\pi^-\}\times\{\pi^-\}$ as a function of $\tau$ at $\sqrt{s}=10.6$ GeV with Collins functions fitted in \cite{Kang:2015msa}. As discussed in Sec.\ \ref{subsec:EECsubset}, we find $|A_{e^+e^-}^{\{\pi^+,\pi^-\}\times\{\pi^+,\pi^-\}}| \ll |A_{e^+e^-}^{\{\pi^+\}\times\{\pi^-\}}|\approx|A_{e^+e^-}^{\{\pi^-\}\times\{\pi^-\}}|$ due to $J^{\perp}_{{\it fav}} \approx -J^{\perp}_{{\it unfav}}$.} \label{fig:asye+e-}
\end{figure}

In Fig.~\ref{fig:asye+e-}, we present the results for $A_{e^+e^-}^{\mathbb{S}_1\times \mathbb{S}_2}(\tau Q^2)$ with different subsets of pions $\mathbb{S}_1\times\mathbb{S}_2=\{\pi^+,\ \pi^-\}\times\{\pi^+,\ \pi^-\}$ (red curve), $\{\pi^+\}\times\{\pi^-\}$ (green curve) and $\{\pi^-\}\times\{\pi^-\}$ (blue curve) at $\sqrt{s}=10.6$ GeV with the error bands given by the uncertainties in the global fitting of the Collins function in~\cite{Kang:2015msa}. Although the subset with $\mathbb{S}_1\times \mathbb{S}_2=\{\pi^+,\ \pi^-\}\times \{\pi^+,\ \pi^-\}$ is suppressed due to the sum rule, one can have sizable asymmetries when restricting to a subset of either positively or negatively charged pions in {{EEC}}.

\subsection{The EEC Collins asymmetry in DIS}\label{sec:phenodis}
To define the asymmetry associated with {{EEC}} for DIS, we begin by rewriting Eq.\eqref{eq:EECDIS3} as
\begin{align}
&\mathrm{{{EEC}}}_{\rm DIS}(\tau,\phi)  = \frac{d\Sigma_{\rm DIS}}{dx_Bdyd\tau d\phi}\nonumber\\
&=\sigma_0\int d^2{\bm q}_T\delta(\tau -\frac{{\bm q}_T^2}{Q^2})\int\frac{db\ b}{2\pi}\bigg\{\mathcal{F}_{UU}+\sin(\phi_{h}+\phi_s)\frac{2(1-y)}{1+(1-y)^2}\mathcal{F}_{UT}^{\sin(\phi_{h}+\phi_s)}+\cdots\bigg\}\,,
\end{align}
where the ellipsis represents other spin-dependent structures given in Eq.~\eqref{eq:EECDIS3} and
\begin{align}
\mathcal{F}_{UU}&=\int d^2{\bm q}_T\delta(\tau -\frac{{\bm q}_T^2}{Q^2})\int\frac{bdb}{2\pi}J_0(bq_T)\,\mathcal{C}\left[\tilde{f}_1\jet\right]\,,\\
\mathcal{F}_{UT}^{\sin(\phi_{h}+\phi_s)}&=\int d^2{\bm q}_T\delta(\tau -\frac{{\bm q}_T^2}{Q^2})\int\frac{bdb}{2\pi}J_1(bq_T)\,\mathcal{C}^\perp\left[\tilde{h}_1\jet^\perp\right]\,,
\end{align}
where the notations $\mathcal{C}^{(\perp)}[\cdots]$ has been defined in Eqs.~\eqref{eq:djq} and~\eqref{eq:dpjq}.
Note that we absorbed $\sqrt{S}$ to define the subtracted versions of TMD PDFs $\tilde{f}_1^{q}$ and $\tilde{h}_1^{q}$ as we did for the EEC jet functions. Then we define the asymmetry as the ratio of the above structure functions
\begin{align}
\mathcal{A}_{\rm DIS}(\tau Q^2)&=\frac{2(1-y)}{1+(1-y)^2}\frac{\mathcal{F}_{UT}^{\sin(\phi_{h}+\phi_s)}}{\mathcal{F}_{UU}}\,.\label{eq:sidis}
\end{align}
As discussed for the $e^+e^-$ case, we can analogously define $\mathcal{A}_{\rm DIS}^{\mathbb{S}}$ for various subsets of hadrons. We use the same parameterization of the twist-3 fragmentation functions discussed around Eq.\ \eqref{eq:kang15_1}-\eqref{eq:kang15_3}. As for the TMD PDFs, we follow the $b^*$ prescription and write them as 
\begin{align}
\tilde{f}_{1,q/p}(x,b, \mu,\zeta) &= 
 \left[\hat{C}_{q\leftarrow i}\otimes f_{1,i/{p}} \right]\left(x,b,\mu_{b_*},\mu_{b_*}^2\right)
e^{-\frac{1}{2}S_{\rm pert}\left(\mu, \mu_{b_*} \right) - S_{\rm NP}^{f_1}\left(b, Q_0, \zeta\right)}\,,
\label{eq:f1tmdevo}
\\
\tilde{h}_{1,q/p}(x,{b},\mu,\zeta) &=\left[\delta C_{q\leftarrow i}\otimes h_{1,i/{p}} \right]\left(x,b,\mu_{b_*},\mu_{b_*}^2\right)
e^{-\frac{1}{2}S_{\rm pert}\left(\mu, \mu_{b_*} \right) - S_{\rm NP}^{h_1}\left(b, Q_0, \zeta\right)}\,,\label{eq:h1tmdevo}
\end{align}
where the coefficients can be found in e.g.~\cite{Kang:2015msa,Boussarie:2023izj} as
\begin{align}
\hat{C}_{q\leftarrow i} (x,b,\mu_{b_*},\mu_{b_*}^2) &=\delta_{qi}\left[\delta(1-x)+\frac{\alpha_{s}}{\pi}\left(-C_{F} \frac{\pi^{2}}{24} \delta\left(1-x\right)+\frac{C_{F}}{2}(1-x)\right)\right]+ \mathcal{O}(\alpha_s^2)\,,\\
\delta C_{q\leftarrow i} (x,b,\mu_{b_*},\mu_{b_*}^2) &= \delta_{qi} \big[\delta(1-x) \big]+ \mathcal{O}(\alpha_s^2)\,,
\end{align}
and the convolution $\otimes$ has been defined in Eq.~\eqref{eq:defconvo}. To parametrize the non-perturbative form factors for TMD PDFs, one has~\cite{Kang:2015msa}
\begin{align}
S_{\rm NP}^{h_1}(b, Q_0,\zeta)&=S_{\mathrm{NP}}^{f_{1}}(b, Q_0,\zeta)=\frac{g_{2}}{2} \ln \left(\frac{b}{b_{*}}\right) \ln \left(\frac{\sqrt\zeta}{Q_{0}}\right)+g_{q} b^{2}\,.
\end{align}
where $g_2=0.84$, $Q_0^2=2.4$ GeV$^2$ and $g_q=0.106$ GeV$^2$. 
For the quark transversity distribution $h_1^{q}\left(x, \mu\right)$, we use the parametrization from\ \cite{Kang:2015msa}.
On the other hand, we use $\mathtt{CT10nlo}$ \cite{Lai:2010vv} for the unpolarized PDFs $f_{1,q/p}(x, \mu)$.

\begin{figure}[h]
\centering
\includegraphics[width=0.54\textwidth]{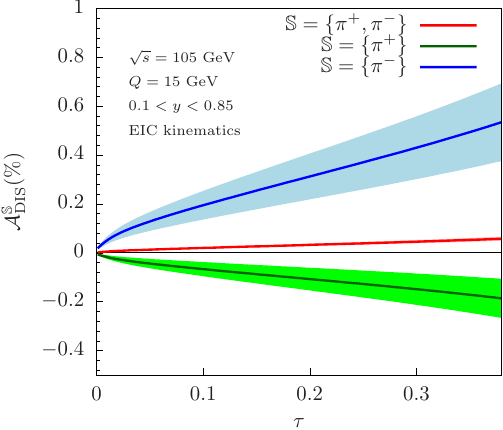}
\caption{$A_{\rm DIS}^{\mathbb{S}}(\tau Q^2)$ for $\mathbb{S}=\{\pi^+\},\ \{\pi^-\}$, and $\{\pi^+,\ \pi^-\}$ using the future EIC kinematics with fittings from~\cite{Kang:2015msa}. Here one also has $|A_{\rm DIS}^{\{\pi^+,\pi^-\}}| \ll |A_{\rm DIS}^{\{\pi^+\}}|$ and $|A_{\rm DIS}^{\{\pi^-\}}|$.}\label{fig:asyDIS}
\end{figure}

In Fig.~\ref{fig:asyDIS}, we present the results for $A_{\rm DIS}^{\mathbb{S}}(\tau Q^2)$ using the future EIC kinematics with different subsets of pions $\mathbb{S}=\{\pi^+,\ \pi^-\}$ (red curve), $\{\pi^+\}$ (green curve) and $\{\pi^-\}$ (blue curve). Here the error bands correspond to the uncertainties in the global extractions~\cite{Kang:2015msa} of transversity TMD PDFs $h_{1,q/p}$ and Collins fragmentation functions $H_1^{\perp q}$. Again, as expected, we find $|A_{\rm DIS}^{\{\pi^+,\pi^-\}}| \ll |A_{\rm DIS}^{\{\pi^+\}}|$ and $|A_{\rm DIS}^{\{\pi^-\}}|$. The asymmetry is overall much smaller than the $e^+e^-$ case due to the cancellation between $h_1^u$ and $h_1^d$ as discussed in the Sec.\ \ref{subsec:EECsubset}. And the asymmetry is measurable when choosing a subset of $\pi^+$ or $\pi^-$.

\subsection{The EEC Sivers asymmetry in DIS}\label{sec:phenosivers}
To define the asymmetry associated with the Sivers asymmetry, we begin by noting that the Eq.\eqref{eq:EECDIS3} has Sivers function dependent part
\begin{align}
&\mathrm{{{EEC}}}_{\rm DIS}(\tau,\phi)  = \frac{d\Sigma_{\rm DIS}}{dx_Bdyd\tau d\phi}\nonumber\\
&=\sigma_0\frac{1}{2}\int d^2{\bm q}_T\delta(\tau -\frac{{\bm q}_T^2}{Q^2})\int\frac{db\ b}{2\pi}\bigg\{\mathcal{F}_{UU}+\sin(\phi_{h}-\phi_s)\frac{2(1-y)}{1+(1-y)^2}\mathcal{F}_{UT}^{\sin(\phi_{h}-\phi_s)}+\cdots\bigg\}\,,
\end{align}
where the ellipsis represents other spin-dependent structures given in Eq.~\eqref{eq:EECDIS3} and
\begin{align}
\mathcal{F}_{UU}=&\int d^2{\bm q}_T\delta(\tau -\frac{{\bm q}_T^2}{Q^2})\int\frac{bdb}{2\pi}J_0(bq_T)\,\mathcal{C}\left[\tilde{f}_1\jet\right]\,,\\
\mathcal{F}_{UT}^{\sin(\phi_{h}-\phi_s)}=&\int d^2{\bm q}_T\delta(\tau -\frac{{\bm q}_T^2}{Q^2})\int\frac{bdb}{2\pi}J_1(bq_T)\,\mathcal{C}\left[\tilde{f}_{1T}^{\perp(1)}\jet\right]\,.
\end{align}
Here the notation $\mathcal{C}[\cdots]$ is provided in Eq.~\eqref{eq:djq}. Note that we absorbed $\sqrt{S}$ to define the subtracted versions of TMD PDFs $\tilde{f}_1$ and $\tilde{f}_{1T}^{\perp(1)}(x,b,\mu,\zeta)$ as we did for the EEC jet functions. Then we define the asymmetry as the ratio of the above structure functions
\begin{align}
\mathcal{A}^{\rm Sivers}_{\rm DIS}(\tau Q^2)&=\frac{2(1-y)}{1+(1-y)^2}\frac{\mathcal{F}_{UT}^{\sin(\phi_{h}-\phi_s)}}{\mathcal{F}_{UU}}\,.\label{eq:sidis}
\end{align}
Now we can study $\mathcal{A}^{\rm Sivers}_{\rm DIS}$ for the final hadron subset $\{\pi^+,\pi^-\}$ as an example. As for the TMD PDFs, we follow the $b^*$ prescription and write them as~\cite{Echevarria:2020hpy} 
\begin{align}
\tilde{f}_{1T,q/p}^{\perp(1)}(x,b,\mu,\zeta) =&-\frac{1}{2M}\int_x^1 \frac{d \hat{x}_1}{\hat{x}_1} \frac{d \hat{x}_2}{\hat{x}_2} \bar{C}_{q \leftarrow i}\left(x / \hat{x}_1, x / \hat{x}_2, b, \mu, \zeta\right) \nnu
&\hspace{1.8cm}\times
T_{F i / p}\left(\hat{x}_1, \hat{x}_2 ; \mu\right)e^{-\frac{1}{2}S_{\rm pert}\left(\mu, \mu_{b_*} \right) - S_{\rm NP}^{f^\perp_{1T}}\left(b, Q_0, \zeta\right)}\,,
\end{align}
where
\begin{align}
\bar{C}_{q \leftarrow i}\left(x_1, x_2, b, \mu_{b_*}, \mu_{b_*}^2\right)=& \delta_{q i}\bigg[ \delta\left(1-x_1\right) \delta\left(1-x_2\right)-\frac{\alpha_s}{2 \pi} \frac{1}{2 N_C} \delta\left(1-x_2 / x_1\right)\left(1-x_1\right) \nnu
&\hspace{0.5cm}-\frac{\alpha_s}{2 \pi}C_F \frac{\pi^2}{12} \delta\left(1-x_1\right) \delta\left(1-x_2\right)\bigg]\,.
\end{align}
And the non-perturbative form factors of TMD PDFs are given by
\begin{align}
S_{\rm NP}^{f_{1T}^\perp}(b,Q_0,Q^2)&=\frac{g_{2}}{2} \ln \left(\frac{b}{b_{*}}\right) \ln \left(\frac{Q}{Q_{0}}\right)+g_{q}^{f_{1T}^\perp} b^{2}\,,
\end{align}
where $g_2=0.84$, $Q_0^2=2.4$ GeV$^2$ and the value of $g_q^{f_{1T}^\perp}$ given by the fitting in ~\cite{Echevarria:2020hpy}, where the Qiu-Sterman function $T_{F\,q/p}(x,x,\mu)$ has the following parametrization form
\begin{align} 
T_{F\,q/p}(x,x,\mu)=N_q \frac{\left(\alpha_q+\beta_q\right)^{\left(\alpha_q+\beta_q\right)}}{\alpha_q^{\alpha_q} \beta_q^{\beta_q}} x^{\alpha_q}(1-x)^{\beta_q}f_{1,q/p}(x,\mu)\,.
\end{align}
where all fitting parameters are presented in \cite{Kang:2015msa}, $f_{1,q/p}$ are the NLO unpolarized PDF set from $\mathtt{HERAPDF20\_NLO\_ALPHAS\_118}$ \cite{H1:2015ubc}, which is applied in the Sivers fitting in ~\cite{Echevarria:2020hpy}.

\begin{figure}[h]
\centering
\includegraphics[width=0.54\textwidth]{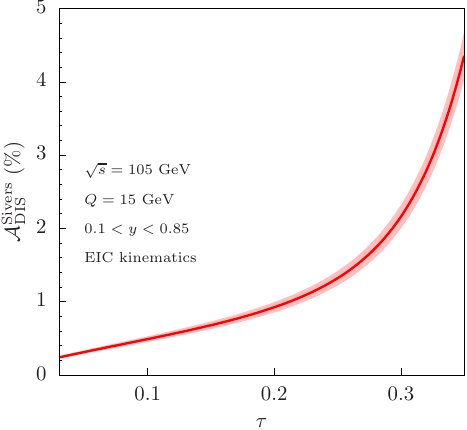}
\caption{$A_{\rm DIS}^{\rm Sivers}(\tau Q^2)$ for $\mathbb{S}=\{\pi^+,\ \pi^-\}$ using the future EIC kinematics with fittings from \cite{Echevarria:2020hpy}. The error band corresponds to the uncertainties in the global extractions of Sivers function in~\cite{Echevarria:2020hpy}. }\label{fig:asySivers2}
\end{figure}

In Fig.~\ref{fig:asySivers2}, we present the results for the Sivers asymmetry $A^{\rm Sivers}_{\rm DIS}(\tau Q^2)$ in terms of EEC jet functions convoluted with TMD PDFs using the parametrization of the twist-3 Qiu-Sterman function fittings from\ \cite{Echevarria:2020hpy}, using the future EIC kinematics with all pions measured in the final state. The magnitude is about a few percents and indicates that this asymmetry is a reasonable measurement for constraining the Sivers function. 

\subsection{The azimuthal asymmetry in DIS using the lab-frame angles}
In this subsection, we provide an azimuthal asymmetry prediction related to the worm-gear function $g_{1T}$, which has been recently extracted from experimental data (see e.g.~\cite{Bhattacharya:2021twu,Horstmann:2022xkk}). In the factorization formalism, one has the EEC defined using the lab frame angular observable $q_*$~\cite{Gao:2022bzi} as we have introduced in Sec.~\ref{sec:labEEC}. As we have shown in Eq.~\eqref{eq:EECDISq20}, the two structure functions for unpolarized and double-spin polarized (worm-gear) incoming proton are given by
\begin{align}
\int db\cos\left(bq_*\right)\mathcal{C}[\tilde{f}_1\jet]=&\int db\cos\left(bq_*\right)\sum_q e_q^2H(Q,\mu)\jet_q(b,\mu,\zeta) \,\tilde{f}_{1,q}(x,b,\mu,\zeta)\,,\label{eq:qsf1}\\
\int db\sin\left(bq_*\right)\mathcal{C}[\tilde{g}^{(1)}_{1T}\jet]=&\int db\sin\left(bq_*\right)\sum_q e_q^2H(Q,\mu)\jet_q(b,\mu,\zeta)\, b\,{M}\,\tilde{g}^{(1)}_{1T,q}(x,b,\mu,\zeta)\,.\label{eq:qsg1t}
\end{align}
For numerical results, using the $b_*$ prescription~\cite{Collins:1984kg} as we have applied in previous sections, we combine TMD evolution with the recent Gaussian fit~\cite{Bhattacharya:2021twu} for the worm-gear function,
\begin{align} 
 g_{1T,q}^{(1)}\left(x, b,\mu,\zeta\right)=& \,g_{1T,q}^{(1)}\left(x, \mu_{b_*}\right) e^{-\frac{1}{2} S_{\mathrm{pert}}\left(\mu, \mu_{b_*}\right)-S_{\mathrm{NP}}^{g_{1T}}(b,Q_0, \zeta)}\,,
 \end{align}
where $S_{\rm pert}(\mu,\mu_{b_*})$ is the same as in Eqs.~\eqref{eq:f1tmdevo} and~\eqref{eq:h1tmdevo}. The parametrization of the non-perturbative Sudakov factor is given by
\begin{align}
S_{\rm NP}^{g_{1T}}(b,Q_0,\zeta)&=\frac{g_2}{2}\ln\left(\frac{b}{b_*}\right)\ln\left(\frac{\sqrt\zeta}{Q_0}\right)+\frac{\langle k_\perp^2\rangle|_{g_{1T}^q}}{4}b^2 \,,\label{eq:snpg1t}
\end{align}
and the fitted function $g_{1 T,q}^{(1)}$ has the following functional form 
\begin{align}
g_{1 T,q}^{(1)}\left(x, Q^2\right)=\frac{N_q}{\widetilde{N}_q} x^{\alpha_q}(1-x)^{\beta_q} f_{1}\left(x, Q^2\right)\,,\label{eq:g1t1}
\end{align}
with the same DGLAP evolution as $f_1(x,Q^2)$ as adopted in the fitting~\cite{Bhattacharya:2021twu}. Here $\langle k_\perp^2\rangle|_{g_{1T}^q}$, $N_q$, $\alpha_q$ and $\beta_q$ are provided in~\cite{Bhattacharya:2021twu}, $\tilde{N}_q$ is defined in terms of the unpolarized collinear PDFs $f_{1}(x,Q_0^2)$ with $Q_0=2$ GeV,
\begin{align}
\tilde{N}_q\equiv\int_0^1 dx x^{\alpha_q+1}(1-x)^{\beta_q}f_{1}(x,Q_0^2)\,.
\end{align}
\begin{figure}[h]
\centering
\includegraphics[width=0.54\textwidth,trim={0 1cm 0 0},clip]{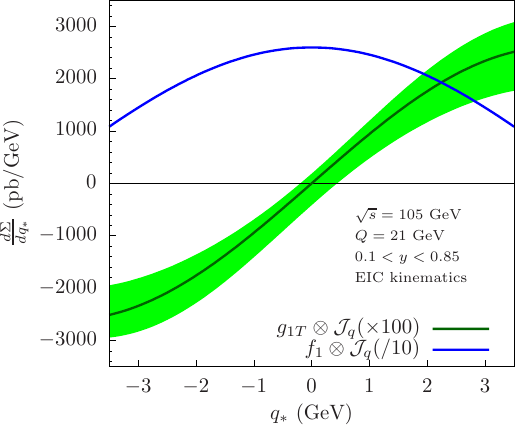}\\
\vspace{-0.15cm}\hspace{0.2cm}\includegraphics[width=0.525\textwidth]{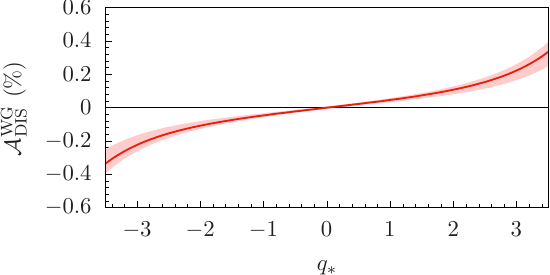}
\\
\caption{The EEC related to unpolarized TMD PDFs $f_1$ (blue) and worm-gear TMD PDFs $g_{1T}$ (green) for all pions in the final state at the EIC kinematics are shown as a function of $q_*$ in the upper panel. The ratio $\mathcal{A}_{\rm DIS}^{\rm WG}$ is shown at the bottom. The error bands in both panels correspond to the uncertainties in the global extractions of worm-gear function in~\cite{Bhattacharya:2021twu}. }\label{fig:h1_ani}
\end{figure}
In the upper panel of Fig.~\ref{fig:h1_ani}, we present the unpolarized structure function in Eq.~\eqref{eq:qsf1} (divided by 10) with a blue curve as a function of $q_*$. We choose the central fit for the unpolarized TMDPDFs, with the collinear PDFs $f_{1}(x,Q^2)$ given by $\mathtt{CT10nlo}$~\cite{Bhattacharya:2021twu}. In the lower panel, we present the structure function depending on the worm-gear functions as given in Eq.~\eqref{eq:qsg1t} (multiplied by 100), plotted as a green curve along with the error band. This error band is again given by the global extraction of the worm-gear functions~\cite{Bhattacharya:2021twu}. In the lower panel of Fig.~\ref{fig:h1_ani}, we use Eq.~\eqref{eq:qsg1t} divided by Eq.~\eqref{eq:qsf1} and plot the ratio $\mathcal{A}^{\rm WG}_{\rm DIS}$ as a function of $q_*$ as well. As given by the extraction in~\cite{Bhattacharya:2021twu}, the worm-gear functions of $u$ quark and $d$ quark have opposite signs and similar magnitudes, and convoluted with unpolarized EEC jet functions, which is always positive. Thus for the production of all pions as final hadrons, the EEC related to worm-gear function are suppressed. This observable can be an inspiration for us to explore new directions for studying 3-dimensional nucleon structures encoded in TMD PDFs.

\section{Conclusion}\label{sec:conclusion}
In summary, we present a comprehensive study of the azimuthal angle dependence of the energy-energy correlators ({{EEC}}) in the back-to-back region to both $e^+e^-$ and deep inelastic scattering (DIS) processes with unpolarized hadron production. Notably, we introduce the Collin-type EEC jet function for the first time in this context. Using the unpolarized and the Collins-type EEC jet function, we demonstrate that all of the leading-twist Transverse-Momentum-Dependent Parton Distribution Functions (TMD PDFs) can be extracted from the DIS process. This finding points towards a new approach for gaining a deeper understanding of the complex structure of nucleons. Furthermore, we introduce a new EEC observable defined only using the lab-frame angle first introduced in~\cite{Gao:2022bzi}, which offers much better experimental resolution.

To illustrate phenomenological applications, we provide predictions for the Collins asymmetry in $e^+e^-$ annihilation and DIS, within the contexts of Belle and Electron-Ion Collider (EIC) kinematics, respectively. Furthermore, we present predictions for the Sivers asymmetry using the Breit frame version of the EEC, as well as the worm-gear asymmetry using the new EEC observable based on lab-frame angles, both with future EIC kinematics. These examples underscore the potential of azimuthal angle dependent EEC as a unique tool for a deeper understanding of nucleon structures.

\acknowledgments
We thank Anjie Gao, Xiaohui Liu, Johannes Michel, Ian Moult, Iain Stewart, Zhiquan Sun, and Feng Yuan for helpful discussions. This work is supported by the National Science Foundation under Grant No.~PHY-1945471 (Z.K. and F.Z.), the U.S. Department of Energy, Office of Science, Office of Nuclear Physics under grant Contract Number DE-SC0011090 (F.Z.), Co-design Center for Quantum Advantage (C2QA) under contract
number DE-SC0012704 (F.Z.), the U.S. Department of Energy under Contract No.~DE-SC0011090 (K.L.), the National Science Foundations of China under Grant No.~12275052 and No.~12147101 and the Shanghai Natural Science Foundation under Grant No.~21ZR1406100 (D.Y.S.).

\appendix
\section{DIS structure functions with hadrons and EEC jets}\label{app1}
In this appendix, we provide the relevant azimuthal-angle-dependent $\bm{q}_T$ factorization in the Breit frame and $q_*$ factorization in the lab frame for the DIS process as well as the structure functions factorized with EEC jet functions.
\subsection{Breit frame adaptations}
As shown in \cite{Gourdin:1972kro,Kotzinian:1994dv,Diehl:2005pc,Bacchetta:2006tn,Anselmino:2008sga,Collins:2011zzd,Kang:2015msa}, the differential cross section for the DIS process up to twist-2 functions is factorized by
\begin{align}
\frac{d\sigma}{dxdydz_hd^2{\bm q}_T}=&\sigma_0\int\frac{bdb}{2\pi}\bigg\{F_{UU}+\cos(2\phi_{h})\frac{2(1-y)}{1+(1-y)^2}F_{UU}^{\cos(2\phi_{h})}\nnu
&+S_{\parallel}\sin(2\phi_{h})\frac{2(1-y)}{1+(1-y)^2}F_{UL}^{\sin(2\phi_{h})}+|{\bm S}_\perp|\bigg[\sin(\phi_{h}-\phi_s)F_{UT}^{\sin(\phi_{h}-\phi_s)}\nnu
&+\sin(\phi_{h}+\phi_s)\frac{2(1-y)}{1+(1-y)^2}F_{UT}^{\sin(\phi_{h}+\phi_s)}\nnu
&+\sin(3\phi_{h}-\phi_s)\frac{2(1-y)}{1+(1-y)^2}F_{UT}^{\sin(3\phi_{h}-\phi_s)}\bigg]\nnu
&+\lambda_e\bigg[S_{\parallel}\frac{y(2-y)}{1+(1-y)^2}F_{LL}+|{\bm S}_\perp|\cos(\phi_{h}-\phi_s)F_{LT}^{\cos(\phi_{h}-\phi_s)}\bigg]\bigg\}\,,\label{dis_cross}
\end{align}
where for structure functions $F_{AB}$, the indices $A$ and $B$ represent the polarization of the incoming electron and incoming proton. $S_{\parallel}$ and $|{\bm S}_\perp|$ are the helicity and transverse spin of the incoming proton, $\lambda_e$ is the helicity of the incoming electron.
The variables $x$ and $y$ represent the longitudinal momentum fraction of the incoming parton inside the beam and the inelasticity variable, respectively, which we will additionally be differential in. The azimuthal angles $\phi_s$ and $\phi_{h}$ are the azimuthal angles of the transverse spin of the beam and the transverse momentum of the produced hadron, respectively. By defining the notation ${C}[\tilde{f}^{(m)}\tilde{D}^{(n)}]$ as
\begin{align}
{C}[\tilde{f}^{(m)}\tilde{D}^{(n)}]=&\sum_q e_q^2 b^{m+n}J_{m+n}(b\,q_T){(M)^m}{(-zM_h)^n}\nnu
&\times\tilde{f}^{(m)}(x,b,\mu,\zeta)\tilde{D}^{(n)}(z,b,\mu,\hat{\zeta})H(Q,\mu)\,,
\end{align}
where $\tilde{f}^{(m)}(x,b,\mu,\zeta)$ and $\tilde{D}^{(n)}(z,b,\mu,\hat{\zeta})$ are the Fourier transform of TMD PDFs and TMD FFs in $b$-space\footnote{To simplify the notations, we apply $n=0$ by default, namely $\tilde{f}_1=\tilde{f}_1^{(0)}$ and $\tilde{D}_1=\tilde{D}_1^{(0)}$.} and defined by~\cite{Boer:2011xd,Kang:2021ffh}
\begin{align}
\tilde{f}^{(m)}(x,b,\mu,\zeta)=&\frac{2\pi m!}{(M^2)^m}\int dp_T p_T\left(\frac{p_T}{b}\right)^m J_m(bp_T)f(x,{\bm p}_T^2,\mu,\zeta)\\
\tilde{D}^{(n)}(z,b,\mu,\hat{\zeta})=&\frac{2\pi n!}{(zM_h^2)^n}\int dk_T k_T\left(\frac{k_T}{b}\right)^n J_n(bk_T)D(z,{\bm k}_T^2,\mu,\hat{\zeta})\,.
\end{align}
Thus one obtains the factorization of the spin-averaged structure function $F_{UU}$ and the all possible spin-dependent structure functions shown in Eq.~\eqref{dis_cross} as
\begin{align}
F_{UU}=&\,C[\tilde{f}_1\tilde{D}_1]\,,\\
F_{UU}^{\cos(2\phi_{h})}=&\,-C[\tilde{h}^{\perp(1)}_1\tilde{H}^{\perp(1)}_1]\,,\\
F_{UL}^{\sin(2\phi_{h})}=&\,-C[\tilde{h}^{\perp(1)}_{1L}\tilde{H}^{\perp(1)}_1]\,,\\
F_{UT}^{\sin(\phi_{h}-\phi_s)}=&\,-C[\tilde{f}^{\perp(1)}_{1T}\tilde{D}_1]\,,\\
F_{UT}^{\sin(\phi_{h}+\phi_s)}=&\,-C[\tilde{h}_1\tilde{H}^{\perp(1)}_1]\,,\\
F_{UT}^{\sin(3\phi_{h}-\phi_s)}=&\,-\frac{1}{4}C[\tilde{h}^{\perp(2)}_{1T}\tilde{H}^{\perp(1)}_1]\,,\\
F_{LL}=&\,C[\tilde{g}_{1L}\tilde{D}_1]\,,\\
F_{LT}^{\cos(\phi_{h}-\phi_s)}=&\,C[\tilde{g}^{\perp(1)}_{1T}\tilde{D}_1]\,,
\end{align}
where we encounter unpolarized and Collins TMD FF from Eq.\ \eqref{eq:diffcskt} again. Here, summation over $q$ includes the anti-quarks as well. Then using the EEC jet functions defined above in Eqs.\ \eqref{eq:unpjet} and\ \eqref{eq:Collinsjet}, namely 
\begin{align}
\jet_q(b,\mu,\zeta)\equiv&\sum_{h}\int_0^1dz\,z\,\tilde{D}_{1,h/q}(z,b,\mu,\zeta)\tag{\ref{eq:unpjet}}\\
\jet_q^{\perp}(b,\mu,\zeta)
\equiv&
\sum_{h}\int_0^1dz\,z\, \tilde{H}_{1,h/q}^{\perp}(z,b,\mu,\zeta)\,.\tag{\ref{eq:Collinsjet}}
\end{align}
Then the azimuthal angle dependent EEC at DIS in Eq.\ \eqref{eq:EECazimDISdef2} can be written as~\footnote{Here the ``subtracted'' TMD PDFs $\tilde{f}_q^{(n)}(x,b,\mu,\zeta)$, EEC jet functions $\jet_q(b,\mu,\zeta)$ and $\jet_q^\perp(b,\mu,\zeta)$ are written as $\tilde{f}^{(n)}_q,\ \jet_q$ and $\jet_q^\perp$ respectively, for simplification of the notations.}

\begin{align}
\frac{d\Sigma_{\rm DIS}}{dxdyd\tau d\phi}=& \sum_{h}\int d^2\bm{q}_T dz_h z_h \frac{d\sigma}{dxdydz_hd^2\bm{q}_T } \delta\left(\tau - \frac{\bm{q}_T^2}{Q^2}\right) \delta(\phi - \phi_{h})\nnu
=&\sigma_0\sum_qe_q^2\int d^2{\bm q}_T\delta(\tau -\frac{{\bm q}_T^2}{Q^2})\delta(\phi-\phi_{h})\int\frac{db\ b}{2\pi}H(Q,\mu)\nnu
&\times \bigg\{J_0(bq_T)\tilde{f}_{1,q}\jet_q+\cos(2\phi_{h})\frac{2(1-y)}{1+(1-y)^2}J_2(bq_T)\frac{b^2}{2}\tilde{h}_{1,q}^{\perp(1)}\jet^\perp_q\nnu
&\quad+S_{\parallel}\sin(2\phi_{h})\frac{2(1-y)}{1+(1-y)^2}J_2(bq_T)\frac{b^2}{2}\tilde{h}_{1L,q}^{\perp(1)}\jet^\perp_q\nnu
&\quad+|{\bm S}_\perp|\bigg[-\sin(\phi_{h}-\phi_s)J_1(bq_T){b}\tilde{f}_{1T,q}^{\perp(1)}\jet_q\nnu
&\quad+\sin(\phi_{h}+\phi_s)\frac{2(1-y)}{1+(1-y)^2}\frac{b}{2}J_1(bq_T)\tilde{h}_{1,q}\jet^\perp_q\nnu
&\quad+\sin(3\phi_{h}-\phi_s)\frac{2(1-y)}{1+(1-y)^2}\frac{M^2b^3}{8}J_3(bq_T)\tilde{h}_{1T,q}^{\perp(2)}\jet^\perp_q\bigg]\nnu
&+\lambda_e\bigg[S_{\parallel}\frac{y(2-y)}{1+(1-y)^2}J_0(bq_T)\tilde{g}_{1L,q}\jet_q
+|{\bm S}_\perp|\cos(\phi_{h}-\phi_s)J_1(bq_T){b}\tilde{g}_{1T,q}^{\perp(1)}\jet_q\bigg]\bigg\}\,,\label{eq:EECDIS2}
\end{align}
where the first term appeared already in\ \cite{Li:2021txc}. 

Unlike the EEC in $e^+e^-$ annihilation, DIS version of the EEC has only one EEC jet function per term as it measures correlation between one outgoing hadron and the beam. By defining the notation $\hat{\mathcal{C}}^{(\perp)}[\tilde{f}^{(m)}\jet^{(\perp)}]$ similar to $\mathcal{C}^{(\perp)}[\tilde{f}^{(m)}\jet^{(\perp)}]$ in Eqs.~\eqref{eq:djq} and~\eqref{eq:djq} but with an extra Bessel function, namely
\begin{align}
\mathcal{C}[\tilde{f}^{(m)}\jet]=&\sum_q e_q^2H(Q,\mu)\jet_q(b,\mu,\zeta) b^{m}{M^m}\tilde{f}^{(m)}_q(x,b,\mu,\zeta)\,,\tag{\ref{eq:djq}}\\
\mathcal{C}^{\perp}[\tilde{f}^{(m)}\jet^{\perp}]=&\sum_q e_q^2H(Q,\mu)\jet_q^\perp(b2,\mu,\zeta)\frac{b^{m+1}}{2}{M^m}{(-M_h)}\tilde{f}^{(m)}_q(x,b,\mu,\zeta) \,,\tag{\ref{eq:dpjq}}\\
\hat{\mathcal{C}}[\tilde{f}^{(m)}\jet]=&\mathcal{C}[\tilde{f}^{(m)}\jet]J_{m}(bq_T)\,,\label{eq:cdef1}\\
\hat{\mathcal{C}}^{\perp}[\tilde{f}^{(m)}\jet^{\perp}]=&\mathcal{C}^{\perp}[\tilde{f}^{(m)}\jet^{\perp}]J_{m+1}(bq_T)\,.\label{eq:cdef2}
\end{align}
Finally, we can define the structure functions $\mathcal{F}_{AB}$ as
\begin{align}
\mathcal{F}_{UU}=&\,\hat{\mathcal{C}}[\tilde{f}_{1}\jet]\,,\\
\mathcal{F}_{UU}^{\cos(2\phi_{h})}=&\,-\hat{\mathcal{C}}^\perp[\tilde{h}^{\perp(1)}_{1}{\jet^\perp}]\,,\\
\mathcal{F}_{UL}^{\sin(2\phi_{h})}=&\,-\hat{\mathcal{C}}^\perp[\tilde{h}^{\perp(1)}_{1L}{\jet^\perp}]\,,\\
\mathcal{F}_{UT}^{\sin(\phi_{h}-\phi_s)}=&\,-\hat{\mathcal{C}}[\tilde{f}^{\perp(1)}_{1T}\jet]\,,\\
\mathcal{F}_{UT}^{\sin(\phi_{h}+\phi_s)}=&\,-\hat{\mathcal{C}}^\perp[\tilde{h}_{1}{\jet^\perp}]\,,\\
\mathcal{F}_{UT}^{\sin(3\phi_{h}-\phi_s)}=&\,-\frac{1}{4}\hat{\mathcal{C}}^\perp[\tilde{h}^{\perp(2)}_{1T}{\jet^\perp}]\,,\\
\mathcal{F}_{LL}=&\,\hat{\mathcal{C}}[\tilde{g}_{1L}\jet]\,,\\
\mathcal{F}_{LT}^{\cos(\phi_{h}-\phi_s)}=&\,\hat{\mathcal{C}}[\tilde{g}^{\perp(1)}_{1T}\jet]\,,
\end{align}
and obtains the azimuthal angle-dependent EEC at DIS as given in Eq.~\eqref{eq:EECDIS3}.

\subsection{EEC using the lab-frame angles}\label{sec:appb}
As introduced in~\cite{Gao:2022bzi}, an optimized observable $q_*$ is defined for probing the transverse momentum of the produced hadron with respect to the photon momentum direction and at leading power $q_*=q_{T}\sin\phi_{h} $. Namely one has
\begin{align}
\frac{d\sigma}{dxdydz_hd{q_*}}=&\int d^2{\bm q}_{T}\delta\left(q_*-q_{T}\sin\phi_{h} \right) \frac{d\sigma}{dxdydz_hd^2{\bm q}_{T}}\,.\label{eq:lab1}
\end{align}
Using the identities 
\begin{align}
J_0\left({bq_{T}}\right):\ &\int d^2{\bm q}_{T}\delta\left(q_*-q_{T}\sin\phi_{h} \right)J_0(bq_{T})=\int q_{T}dq_{T}\frac{d\phi_{h}}{\sin\phi_{h}}\delta\left(\frac{q_*}{\sin\phi_{h}}-{q_{T}}\right)J_0\left({bq_{T}}\right)\nnu
=&\int \frac{\phi_{h}}{\sin^2\phi_{h}}q_*\Theta\left(\frac{q_*}{\sin\phi_{h}}\right)J_0\left(\frac{bq_*}{\sin\phi_{h}}\right)=\,\frac{2\cos\left(bq_*\right)}{b}\,,\label{eq:qs_int_1}\\
J_1\left({bq_{T}}\right):\ &\int d^2{\bm q}_{T}\delta\left(q_*-q_T\sin\phi_{h}\right)\sin\left(\phi_{h}\right)J_1({bq_{T}})\nnu
=&\int \frac{\phi_{h}}{\sin^2\phi_{h}}q_*\Theta\left(\frac{q_*}{\sin\phi_{h}}\right)\sin\left(\phi_{h}\right)J_1\left(\frac{bq_*}{\sin\phi_{h}}\right)=\,\frac{2\sin\left(bq_*\right)}{b}\,,\\
&\int d^2{\bm q}_T\delta\left(q_*-q_T\sin\phi_{h}\right)\cos\left(\phi_{h}\right)J_1(bq_T)=0\,,\\
J_2\left({bq_{T}}\right):\ &\int d^2{\bm q}_T\delta\left(q_*-q_T\sin\phi_{h}\right)\sin\left(2\phi_{h}\right)J_2({bq_{T}})=0\,,\label{eq:h1lp}\\
&\int d^2{\bm q}_T\delta\left(q_*-q_T\sin\phi_{h}\right)\cos\left(2\phi_{h}\right)J_2({bq_{T}})\nnu
=&\int \frac{\phi_{h}}{\sin^2\phi_{h}}q_*\Theta\left(\frac{q_*}{\sin\phi_{h}}\right)\cos\left(2\phi_{h}\right)J_2\left(\frac{bq_*}{\sin\phi_{h}}\right)=\,\frac{2\cos\left(bq_*\right)}{b}\,,\\
J_3\left({bq_{T}}\right):\ &\int d^2{\bm q}_T\delta\left(q_*-q_T\sin\phi_{h}\right)\sin\left(3\phi_{h}\right)J_3(bq_T)\nnu
=&\int \frac{\phi_{h}}{\sin^2\phi_{h}}q_*\Theta\left(\frac{q_*}{\sin\phi_{h}}\right)\sin\left(3\phi_{h}\right)J_3\left(\frac{bq_*}{\sin\phi_{h}}\right)=\,\frac{2\sin\left(bq_*\right)}{b}\,,\\
&\int d^2{\bm q}_T\delta\left(q_*-q_T\sin\phi_{h}\right)\cos\left(3\phi_{h}\right)J_3(bq_T)=0\,,\label{eq:qs_int_2}
\end{align}
following the DIS differential cross section introduced in App.~\ref{app1} and Eq.~\eqref{eq:lab1}, one obtains
\begin{align}
\frac{d\sigma}{dxdydz_hd{q_*}}=&\sigma_0\int db\sum_q e_q^2H(Q,\mu)S({\bm b}^2,\mu,\nu)\bigg\{\cos\left(bq_*\right)\bigg[\tilde{f}_{1}\tilde{D}_1\nnu
&+\frac{2(1-y)}{1+(1-y)^2}zMM_hb^2\tilde{h}_{1}^{\perp (1)}\tilde{H}_1^{\perp (1)}+\lambda_eS_{\parallel}\frac{y(2-y)}{1+(1-y)^2}\tilde{g}_{1L}\tilde{D}_1\bigg]\nnu
&+|{\bm S}_\perp|\sin\left(bq_*\right)\cos\phi_s\bigg[-Mb\tilde{f}_{1T}^{\perp(1)}\tilde{D}_1+\frac{2(1-y)}{1+(1-y)^2}zM_hb\tilde{h}_{1}\tilde{H}_1^{\perp (1)}\nnu
&+\frac{2(1-y)}{1+(1-y)^2}\frac{zM^2M_hb^3}{4}\tilde{h}_{1T}^{\perp (2)}\tilde{H}_1^{\perp (1)}\bigg]+\lambda_e|{\bm S}_\perp|\sin\left(bq_*\right)\sin\phi_sM\tilde{g}_{1T}^{(1)}\tilde{D}_1\bigg\}\,.
\end{align}
Note that as indicated in Eq.~\eqref{eq:h1lp}, the contribution from $h_{1L}^\perp$ will vanish when measuring $q_*$. Now we define the EEC with $q_*$ measured as
\begin{align}
\frac{d\Sigma_{\rm DIS}}{dxdydq_*}=& \sum_{h}\int d^2\bm{q}_T dz_h z_h \frac{d\sigma}{dxdy dz_hd^2\bm{q}_T} \delta(q_* - q_T\sin\phi_{h}) \nnu
=&\sigma_0\int db\bigg\{\cos\left(bq_*\right)\bigg[\mathcal{C}[\tilde{f}_1\jet_q]\nnu
&+\frac{2(1-y)}{1+(1-y)^2}\mathcal{C}^{\perp}[\tilde{h}^{\perp(1)}\jet_q^{\perp}]+\lambda_eS_{\parallel}\frac{y(2-y)}{1+(1-y)^2}\mathcal{C}[\tilde{g}_{1L}\jet_q]\bigg]\nnu
&+|{\bm S}_\perp|\sin\left(bq_*\right)\cos\phi_s\bigg[-\mathcal{C}[\tilde{f}^{\perp(1)}_{1T}\jet_q]+\frac{2(1-y)}{1+(1-y)^2}\mathcal{C}^{\perp}[\tilde{h}_1\jet_q^{\perp}]\nnu
&+\frac{2(1-y)}{1+(1-y)^2}\frac{1}{4}\mathcal{C}^{\perp}[\tilde{h}^{\perp(2)}_{1T}\jet_q^{\perp}]\bigg]+\lambda_e|{\bm S}_\perp|\sin\left(bq_*\right)\sin\phi_s\mathcal{C}[\tilde{g}^{(1)}_{1T}\jet_q]\bigg\}\,,\label{eq:EECDISq22}
\end{align}
where
\begin{align}
\mathcal{C}[\tilde{f}^{(m)}\jet]=&\sum_q e_q^2H(Q,\mu)\jet_q(b,\mu,\zeta) b^{m}{M^m}\tilde{f}^{(m)}_q(x,b,\mu,\zeta)\,,\tag{\ref{eq:djq}}\\
\mathcal{C}^{\perp}[\tilde{f}^{(m)}\jet^{\perp}]=&\sum_q e_q^2H(Q,\mu)\jet_q^\perp(b,\mu,\zeta)\frac{b^{m+1}}{2}{M^m}{(-M_h)}\tilde{f}^{(m)}_q(x,b,\mu,\zeta) \,.\tag{\ref{eq:dpjq}}
\end{align}
Compared to $\hat{\mathcal{C}}[\cdots]$ defined in Eqs.~\eqref{eq:cdef1} and~\eqref{eq:cdef2}, the notation here $\mathcal{C}[\cdots]$ is not equipped with a Bessel function, since $J_{m(+1)}(bq_T)$ has been integrated out in Eqs.~\eqref{eq:qs_int_1}-\eqref{eq:qs_int_2} for measuring $q_*$.

Upon closer inspection of Eq.~\eqref{eq:EECDISq22} and Eq.(10) in~\cite{Gao:2022bzi}, the disparity in sign arises from different definitions of the $m$-th order TMD PDFs in $b$-space. Nonetheless, one can successfully harmonize the two works by appropriately factoring in $(-1)^m$ for each structure function in Eq.~\eqref{eq:EECDISq22} that corresponds to $\tilde{f}^{(m)}$, and thereby cancel the divergence in convention.

\bibliographystyle{JHEP}
\bibliography{poleec}

\end{document}